\documentclass[10pt, twocolumn, twoside]{IEEEtran}

\usepackage{cite}
\usepackage{enumerate}
\usepackage{graphicx}
\usepackage[cmex10]{amsmath} 
\usepackage{amssymb}
\usepackage{mathtools}
\usepackage{xspace}
\usepackage{subfigure}
\usepackage[lined,linesnumbered,ruled,commentsnumbered]{algorithm2e} 
\usepackage{colonequals}
\usepackage[mathscr]{euscript}
\usepackage{fixltx2e}    
\usepackage{amsthm}
\usepackage{mathrsfs}

\usepackage{epsfig}
\usepackage{epstopdf}

\usepackage{tikz}
\usepackage{pgfplots}

\usetikzlibrary{arrows,shapes,backgrounds,plotmarks,positioning}

\pgfplotsset{compat=newest}                         
\pgfplotsset{plot coordinates/math parser=false}
\newlength\figureheight
\newlength\figurewidth

\newtheorem{theorem}{Theorem}
\newtheorem{lemma}[theorem]{Lemma}
\newtheorem{definition}[theorem]{Definition}

\newtheorem{example}[theorem]{Example}
\newtheorem{remark}[theorem]{Remark}

\newcommand{\RZ}[1]{\mathsf{Z}_{#1}}
\newcommand{\ARZ}[1]{\mathcal{Z}_{#1}} 
\newcommand{\RW}[1]{\mathsf{W}_{#1}}
\newcommand{\RCO}{R^*}
\newcommand{\RR}{\mathscr{R}}
\newcommand{\RRCO}{\mathscr{R}^*}
\newcommand{\MI}{\Lambda}

\newcommand{\G}[2]{\Omega(#1,#2)}

\newcommand{\Set}[1]{\{#1\}}

\newcommand{\Pat}{\mathcal{P}}

\newcommand{\rv}{\mathbf{r}} 
\newcommand{\rvS}{\hat{\mathbf{r}}} 
\newcommand{\rS}{\hat{r}} 
\newcommand{\rvE}{\mathbf{r}^*} 
\newcommand{\rE}{r^*} 
\newcommand{\wv}{\mathbf{w}}         
\newcommand{\One}{\mathbf{1}}      

\newcommand{\Fu}[1]{f_{#1}}
\newcommand{\FuHat}[1]{\hat{f}_{#1}}

\newcommand{\Real}{\mathbb{R}}
\newcommand{\RealP}{\mathbb{R}_{+}}    
\newcommand{\RealPP}{\mathbb{R}_{++}}    
\newcommand{\Z}{\mathbb{Z}}            
\newcommand{\ZP}{\mathbb{Z}_{+}}    
\newcommand{\ZPP}{\mathbb{Z}_{++}}    
\newcommand{\F}{\mathbb{F}}            
\newcommand{\Q}{\mathbb{Q}}            

\newcommand{\SFM}{\text{SFM}}

\newcommand{\EX}{\text{EX}}

\newcommand{\figref}[1]{Fig.~\ref{#1}}

\newcommand{\dep}{\mathrm{dep}}

\newcommand{\argmin}{\text{argmin}}
\newcommand{\argmax}{\text{argmax}}

\newcommand{\NormOne}[1]{\lVert #1 \rVert_1}

\begin{document}

\title{Attaining Fairness in Communication for Omniscience}

\author{Ni~Ding,~\IEEEmembership{Member,~IEEE}, Parastoo~Sadeghi,~\IEEEmembership{Senior Member,~IEEE}, David~Smith,~\IEEEmembership{Member,~IEEE}, and Thierry~Rakotoarivelo,~\IEEEmembership{Member,~IEEE}

\thanks{Some initial results of this paper have been published in \cite{Ding2016ISIT,Ding2015ICT,Ding2015Game,Ding2018ISITShapley}.}
\thanks{Ni Ding, David Smith and Thierry Rakotoarivelo (email: $\{$ni.ding, david.smith, thierry.rakotoarivelo$\}$@data61.csiro.au) are with Data61, 5/13 Garden Street, Eveleigh, NSW 2015.}
\thanks{Parastoo Sadeghi (email: $\{$parastoo.sadeghi$\}$@anu.edu.au) is with Research School of Electrical, Energy and Materials Engineering (RSEEME), The Australian National University, ACT, 2601. }
}

\maketitle

\begin{abstract}
    This paper studies how to attain fairness in communication for omniscience, where a set of users exchange their observations of a discrete multiple random source to attain omniscience---the state that all users recover the entire source.
    The optimal rate region containing all source coding rate vectors that achieve the omniscience with the minimum sum rate is shown to coincide with the core (the solution set) of a coalitional game.
    Two game-theoretic fairness solutions are studied: the Shapley value and the egalitarian solution. It is shown that the Shapley value assigns each user the source coding rate measured by his/her remaining information of the multiple source given the common randomness that is shared by all users, while the egalitarian solution simply distributes the rates as evenly as possible in the core.
    %
    To avoid the exponentially growing complexity of obtaining the Shapley value, a polynomial-time approximation method is proposed by utilizing the fact that the Shapley value is the mean value over all extreme points in the core.
    In addition, a steepest descent algorithm is proposed which converges in polynomial time to the fractional egalitarian solution in the core that can be implemented by network coding schemes.
    Finally, it is shown that the game can be decomposed into subgames so that both the Shapley value and the egalitarian solution can be obtained within each subgame in a distributed manner with reduced complexity.
\end{abstract}

\begin{IEEEkeywords}
Coalitional game, communication for omniscience, fairness, submodularity.
\end{IEEEkeywords}

\section{introduction}

The communication for omniscience (CO) problem is formulated in \cite{Csiszar2004}. It is assumed that there are a finite number of users in a system that are indexed by the set $V$. Each user $i \in V$ observes a distinct component $\RZ{i}$ of a discrete multiple random source $\RZ{V} = (\RZ{i} \colon i \in V)$ in private. The users are allowed to exchange their observations over public authenticated broadcast channels so as to attain \textit{omniscience}, the state where each user recovers the observation sequence of the entire source $\RZ{V}$.
Originally, the CO problem was studied in \cite{Csiszar2004} due to its dual relationship with the multi-terminal secret capacity \cite[Theorem~1]{Csiszar2004}. More recently, the system model was also cast into the \textit{coded cooperative data exchange (CCDE)} problem \cite{Roua2010,Court2010,Court2010M} in which the users are mobile clients broadcasting linear combinations of packets over noiseless peer-to-peer (P2P) wireless channels and the communication rates are restricted to be integral.

One main optimization problem that arises in CO is how to minimize the overall source coding rate to attain omniscience. We call it the \emph{minimum sum-rate problem} and denote the value of the minimum sum-rate by $\RCO$. By utilizing submodular function minimization (SFM) techniques, the value of $\RCO$, as well as an optimal rate vector are determined in $O(|V|^2 \cdot \SFM(|V|)$ time in \cite{Ding2018IT} for the asymptotic model where the communication rates are real-valued.\footnote{In an asymptotic model, the observation sequence is assumed to be infinitely long. The CCDE corresponds to the finite linear source model, an example of the non-asymptotic model. In the non-asymptotic model, each user only obtains a finite length of observations and the broadcasts are integer number of linear combinations of observations \cite[Section~II]{Ding2018IT}.} Here, $\SFM(|V|)$ is the complexity of a SFM algorithm and is polynomial \cite[Chapter~VI]{Fujishige2005}.
For solving the minimum sum-rate problem in CCDE, the authors in \cite{MiloIT2016,CourtIT2014} proposed deterministic algorithms, which also complete in $O(|V|^2 \cdot \SFM(|V|))$ time. In addition, the algorithm proposed in \cite{SprintRand2010} determines the minimum sum-rate in CCDE more efficiently by simulating the communications based on the random linear network coding scheme \cite{NIF2000,RLNC2006}.

While existing algorithms in \cite{Ding2018IT,MiloIT2016,CourtIT2014,SprintRand2010} only determine one optimal rate vector, it is shown in \cite[Section ~III-B]{Ding2018IT} that the optimal rate region is not a singleton in general. So, it is natural to consider how to choose an optimal rate vector that also attains fairness, in particular when the intention is to promote the mobile clients' cooperation in CCDE or even out the battery usage in a wireless sensor network (WSN).
The problem of how to attain fairness has been previously considered in \cite{MiloFair2012,Taj2011} for CCDE. In \cite{Taj2011}, a multi-layer acyclic graph is proposed, based on which, a constrained quadratic programming is formulated to determine the Jain's fairness solution \cite{Jain1984}.
The algorithm proposed in \cite{MiloFair2012} is a greedy approach, where, in each iteration, a unit rate is assigned to the user that optimizes a fairness measure, so that the resulting solution converges to a fair and integer-valued optimal rate vector.
In fact, these two methods both aim at determining the integral egalitarian solution in the optimal rate region.\footnote{The fair solutions in \cite{MiloFair2012,Taj2011} coincide with the egalitarian solution \cite{Dutta1989Egalitarian} in coalitional game theory due to the equivalence between the submodular base polyhedron and the optimal rate region \cite[Section~III-B]{Ding2018IT}, both of which, as will be shown in Section~\ref{sec:Game} in this paper, coincide with the core of a coalitional game. } However, neither of them applies to systems where the communication rates are non-integral, e.g., the asymptotic model or where packet splitting (and hence fractional transmission rates) is allowed in CCDE.

The main purpose of this paper is to study how to attain fairness in the optimal rate region for the CO problem, where the broadcast rates are not constrained to be integer-valued. We start the study by showing the equivalence between the optimal rate region and the core (the solution set) of a coalitional game.
It is shown that the game formulates the multi-terminal source coding problem for attaining the omniscience of multiple source $\RZ{V}$ by the minimum sum-rate $\RCO$.
We then consider two fair solutions proposed in coalitional game theory: the Shapley value \cite{Shapley1953Value} and the egalitarian solution \cite{Dutta1989Egalitarian}. The Shapley value assigns each user the expected marginal remaining randomness given the common information shared by all users, whereas the egalitarian solution simply tries to even out the source coding rates in the optimal region. To alleviate the exponentially growing complexity of obtaining the Shapley value with problem size, we show how to approximate it in polynomial time. We also propose a steepest descent algorithm (SDA) for searching a fractional egalitarian solution that can be implemented by packet splitting in CCDE.
Finally, we show that the game can be decomposed by the fundamental partition $\Pat^*$ into subgames,\footnote{The fundamental partition $\Pat^*$ is an optimizer that determines the minimum sum-rate $\RCO$ \cite{Ding2018IT}. See also Section~\ref{sec:review}.}
each of which can attain fairness, either being the Shapley value or the egalitarian solution, on its own. This decomposition leads to a distributed computation method for attaining fairness and also reduces the complexity.

\subsection{Summary of Main Results}

Our main results are summarized as follows:

1) We formulate the problem of attaining the omniscience with the minimum sum-rate $\RCO$ by a coalitional game model, where the characteristic cost function $\FuHat{\RCO} (X)$ for all $X \subseteq V$ quantifies the remaining randomness in $\RZ{X}$ given the common randomness $\MI = H(V) - \RCO$ that is shared by all users in $V$.\footnote{The game model is closely related to the dual relationship \cite[Theorem~1]{Csiszar2004,Chan2008tight}: $\RCO = H(V) - \MI$, where $H(V)$ is the entropy of $\RZ{V}$ and $\MI$ is the common randomness that is shared by all the users in $V$ \cite{Prakash2016,ChanMMI}. The interpretation is that attaining omniscience by the minimum sum-rate $\RCO$ is equivalent to determining how to let the users encode the remaining randomness in $\RZ{V}$ given the common randomness $\MI$.}
Since $\FuHat{\RCO}$ specifies the source coding rate/cost upper bound to each user subset $X \subseteq V$, we show that (a) the core of the game coincides with the optimal rate region containing all the solutions to the minimum sum-rate problem and (b) the game describes the users' cooperation when they jointly encode the remaining randomness in $\RZ{V}$ to reach the omniscience with the sum-rate exactly equal to $\RCO$.

2) Based on the game model, we introduce the first fairness solution, the Shapley value \cite{Shapley1953Value}. We show that the Shapley value assigns each user the expected marginal cost $\FuHat{\RCO}(X \cup \Set{i}) - \FuHat{\RCO} (X)$ over all $X \subseteq V \setminus \Set{i}$. This solution is fair in that it penalizes each user according to the source coding rate he/she incurred in CO. While the complexity of obtaining the exact Shapley value is exponentially growing in $|V|$, we show that the Shapley value is the mean over all extreme points in the core. By randomly generating an extreme point set of a desired size, we are able to obtain an approximation of the Shapley value in polynomial time. 

3) The egalitarian solution \cite{Dutta1989Egalitarian} aims to equalize the rate/cost allocation in the optimal rate region regardless of the marginal costs. This solution is more suitable for those systems with equally privileged users, e.g., CCDE and WSN. While there exist polynomial-time algorithms in the literature, e.g., \cite{Nagano2012Lex,Nagano2013}, that determine a real-valued egalitarian solution, we propose a steepest descent algorithm (SDA) for searching a fractional egalitarian solution that can be implemented in CCDE by splitting each packet into $|\Pat^*| - 1$ chunks.
Based on an optimality criterion for the egalitarian solution stating that the local optimum implies the global optimum, we show that the estimation sequence generated by the SDA converges to the fractional egalitarian solution in $O( |\Pat^*| \cdot L(V) \cdot |V| \cdot \SFM(|V|))$ time, where $L(V)$ is the maximum $\ell_1$-norm over all pairs of points in the optimal rate region.
In addition, the steepest direction in each iteration of SDA can be computed in a distributed manner.

4) We show that the coalitional game model can be decomposed by the fundamental partition $\Pat^*$: the users in each subset $C \in \Pat^*$ form a subgame with the characteristic cost function $\FuHat{\RCO}(X)$ for all $X \subseteq C$. This decomposition is due to the mutual independence between $\RZ{C}$ and $\RZ{C'}$ for any two distinct subsets $C,C' \in \Pat^*$ given the common randomness $\MI$.
For attaining fairness in the optimal rate region, it suffices to let the users within each subgame $C \in \Pat^*$ decide how to allocate the source coding rates fairly. This allows a decomposition method to reach fair solutions: the fusion of the Shapley values and the egalitarian solutions over all subgames constructs the Shapley value and the egalitarian solution, respectively, of the entire game, which not only reduces complexity, but also allows parallel computation.

\subsection{Organization}

The rest of the paper is organized as follows. The system model is described in Section~\ref{sec:system}, where we also review existing results on the minimum sum-rate problem. In Section~\ref{sec:Game}, we formulate the coalitional game model and show that it can be decomposed by the fundamental partition $\Pat^*$. In Section~\ref{sec:Shapley}, we show how to attain fairness in the optimal rate region by the Shapley value and discuss how to approximate it to avoid the exponentially growing complexity. In Section~\ref{sec:Egalitarian}, we propose the SDA algorithm for searching the fractional egalitarian solution. In both Sections~\ref{sec:Shapley} and \ref{sec:Egalitarian}, we also present methods to obtain the Shapley value and egalitarian solution by the decomposition method.

\section{Communication for Omniscience}
\label{sec:system}

Let $V$ with $|V|>1$ be a finite set that indexes the terminals in a discrete memoryless multiple source $\RZ{V}=(\RZ{i}:i\in V)$. Each component $\RZ{i}$ is a discrete random variable that takes its values in the finite alphabet $\ARZ{i}$ according to the joint probability mass function $P_{\RZ{V}}$. Let there be $|V|$ users. Each user $i \in V$ observes an i.i.d.\ $n$-sequence $\RZ{i}^n$ of the component $\RZ{i}$ in private.
The users are allowed to exchange compressed versions of their observations over noiseless broadcast channels. The purpose is to attain \textit{omniscience}, the state where all users recover the observation sequence $\RZ{V}^n$. This problem is called \textit{communication for omniscience (CO)} \cite{Csiszar2004}.\footnote{The CO problem was originally formulated in \cite{Csiszar2004} based on a study on the secret capacity in a more general setting where a set of users $A \subseteq V$ serve as helpers that assist the active users in generating the secret key. The CO problem considered in this paper is the case when $A = V$.}

\subsection{Minimum Sum-rate and Optimal Rate Region}
\label{sec:review}

For $X \subseteq V$, let $H(X)$ be the amount of randomness in $\RZ{X}$ measured by Shannon entropy \cite{Cover2012ITBook}. For a \emph{(source coding) rate vector} $\rv_V=(r_i:i\in V)$, each dimension $r_i$ denotes the code rate at which user $i$ encodes his/her observation $\RZ{i}^{n}$.
Let $r \colon 2^V \mapsto \RealP$ be the \textit{sum-rate function} associated with $\rv_V$ such that
$$ r(X)=\sum_{i\in X} r_i, \quad \forall X \subseteq V, $$
with the convention $r(\emptyset)=0$.
Here, $r(X)$ denotes the rates at which the users in $X$ jointly encode $\RZ{X}^n$.
A source coding rate vector $\rv_V$ at which omniscience is attainable satisfies the Slepian-Wolf (SW) constraints $r(X) \geq H(X|V \setminus X),\forall X \subsetneq V$ \cite{Csiszar2004}. The \emph{achievable rate region} is
\begin{equation} \label{eq:RateReg}
    \RR(V)=\Set{ \rv_V \in \Real^{|V|} \colon r(X) \geq H(X|V\setminus X),\forall X \subsetneq V }.
\end{equation}

The fundamental problem concerning the efficiency in CO is to minimize the sum-rate for attaining omniscience
\begin{equation} \label{eq:MinSumRateACO}
    \RCO = \min\Set{r(V) \colon \rv_V\in \RR(V)}.
\end{equation}
This \emph{minimum sum-rate problem} has been studied and solved efficiently in \cite{Ding2016NetCod,Ding2018IT} without dealing with the exponentially large number of constraints in the linear programming \eqref{eq:MinSumRateACO}. We review some results in \cite{Ding2018IT} as follows. They will be used in Section~\ref{sec:Game} to formulate the game model.

For sum-rate $\alpha \in \RealP$, define
$$ \Fu{\alpha}(X)=\begin{cases} 0 & X = \emptyset \\ \alpha - H(V \setminus X | X) & X \neq \emptyset \end{cases}. $$
Let $\Pi(V)$ be the set containing all partitions of $V$. The \textit{Dilworth truncation} of $\Fu{\alpha}$ is $\FuHat{\alpha}(X) = \min_{\Pat \in \Pi(X)}\sum_{C \in \Pat} \Fu{\alpha}(C)$ for all $X \subseteq V $ \cite{Dilworth1944}.
It is shown in \cite[Theorem 4 and Corollary 46]{Ding2018IT} that
\begin{equation}\label{eq:MinSuumRateDilworth}
    \RCO = \min \Set{\alpha \colon \Fu{\alpha}(V) = \FuHat{\alpha}(V)}.
\end{equation}
The \emph{optimal rate region} $\RRCO(V)$ that contains all achievable rate vectors $\rv_V$ with sum-rate $r(V) = \RCO$ coincides with $B(\FuHat{\RCO})$, the \emph{base polyhedron} of $\FuHat{\RCO}$ \cite[Section 2.3]{Fujishige2005} \cite[Definition 9.7.1]{Narayanan1997Book}:
\begin{equation} \label{eq:Equivalence}
    \begin{aligned}
        \RRCO(V) &= \Set{ \rv_V \in \RR(V) \colon r(V) = \RCO} \\
                 &= \Set{\rv_V \in P(\FuHat{\RCO}) \colon r(V) = \FuHat{\RCO}(V) = \RCO} \\
                 &= B(\FuHat{\RCO}),
    \end{aligned}
\end{equation}
where $P(\FuHat{\RCO}) = \Set{\rv_V\in\Real^{|V|} \colon r(X) \leq \FuHat{\RCO}(X),\forall X \subseteq V}$ is the \emph{polyhedron} of $\FuHat{\RCO}$, which coincides with $P(\Fu{\RCO}) = \Set{\rv_V\in\Real^{|V|} \colon r(X) \leq \Fu{\RCO}(X),\forall X \subseteq V}$ \cite[Theorems 2.5(i) and 2.6(i)]{Fujishige2005}. Here, the polyhedron $P(\Fu{\RCO})$ is induced by the SW constraints: the inequality $r(X) \geq H(X | V \setminus X)$ in \eqref{eq:RateReg} is converted to $r(V \setminus X) \leq \RCO - H(X | V \setminus X) $ under the constraint $r(V) = \RCO$ in $ B(\FuHat{\RCO})$.

Problem~\eqref{eq:MinSuumRateDilworth} can be solved in $O(|V|^2 \cdot \SFM(|V|) )$ time by the modified decomposition algorithm (MDA) proposed in \cite[Section V-A]{Ding2018IT},\footnote{The efficiency of the MDA algorithm relies on the submodularity of the entropy function $H$. $\SFM(|V|)$ denotes the complexity of solving a submodular function. See Appendix~\ref{app:Prelim} for the definition of the submodularity and a brief note on $\SFM(|V|)$. }
which also returns an optimal rate vector in $\RRCO(V)$. Let $\Pat^*$ be the finest minimizer that determines the Dilworth truncation
\begin{equation} \label{eq:Dilworth}
    \FuHat{\RCO}(V) = \min_{\Pat \in \Pi(V)}\sum_{C \in \Pat} \Fu{\RCO}(C).
\end{equation}
We call $\Pat^*$ the \textit{fundamental partition}, which is also returned by the MDA algorithm.

\subsection{Fairness}
While the optimal rate region $\RRCO(V)$ is not necessarily a singleton, the MDA algorithm, as well as \cite[Algorithm~3]{MiloIT2016} \cite[Appendix~F]{CourtIT2014} for solving the minimum sum-rate problem in CCDE determine an extreme point (a vertex) in $\RRCO(V)$, as illustrated in the following example.

\begin{figure}[tpb]
	\centering
    \scalebox{0.95}{\begin{tikzpicture}

\draw (-3.2,0.3) rectangle (-1.8,-0.3);
\node at (-2.5,0) {user $1$};
\draw (-2.5,0.3)--(-2.5,1)--(-2.2,1)--(-2.5,0.8)--(-2.8,1)--(-2.5,1);
\node at (-2.5,-0.5) {\scriptsize \textcolor{blue}{$\RZ{1} = (\RW{b},\RW{c},\RW{d},\RW{h},\RW{i})$}};

\draw (3.2,0.3) rectangle (1.8,-0.3);
\node at (2.5,0) {user $4$};
\draw (2.5,0.3)--(2.5,1)--(2.2,1)--(2.5,0.8)--(2.8,1)--(2.5,1);
\node at (2.5,-0.5) {\scriptsize \textcolor{blue}{$\RZ{4} = (\RW{a},\RW{b},\RW{c},\RW{d},\RW{f},\RW{g},\RW{i},\RW{j})$}};

\draw (-0.7,1.6) rectangle (0.7,1);
\node at (0,1.3){user $5$};
\draw (0,1.6)--(0,2.3)--(0.3,2.3)--(0,2.1)--(-0.3,2.3)--(0,2.3);
\node at (0,0.8) {\scriptsize \textcolor{blue}{$\RZ{5} = (\RW{a},\RW{b},\RW{c},\RW{f},\RW{i},\RW{j})$}};

\draw (-2.5,-1.7) rectangle (-1.1,-2.3);
\node at (-1.8,-2) {user $2$};
\draw (-1.8,-1.7)--(-1.8,-1)--(-1.5,-1)--(-1.8,-1.2)--(-2.1,-1)--(-1.8,-1);
\node at (-1.8,-2.5) {\scriptsize \textcolor{blue}{$\RZ{2} = (\RW{e},\RW{f},\RW{h},\RW{i})$}};

\draw (2.5,-1.7) rectangle (1.1,-2.3);
\node at (1.8,-2) {user $3$};
\draw (1.8,-1.7)--(1.8,-1)--(1.5,-1)--(1.8,-1.2)--(2.1,-1)--(1.8,-1);
\node at (1.8,-2.5) {\scriptsize \textcolor{blue}{$\RZ{3} = (\RW{b},\RW{c},\RW{e},\RW{j})$}};

\end{tikzpicture}}
	\caption{The $5$-user system with $V = \Set{1,\dotsc,5}$ in Example~\ref{ex:main}. The users encode and broadcast $\RZ{i}$s so as to attain the omniscience of the source $\RZ{V}$. In the corresponding CCDE problem, each $\RW{j}$ denotes a packet that belongs to a field $\F_q$ and each user $i \in V$ broadcasts linear combinations of $\RZ{i}$ to help others recover all packets in $\RZ{V}$. }
	\label{fig:CDESystem}
\end{figure}
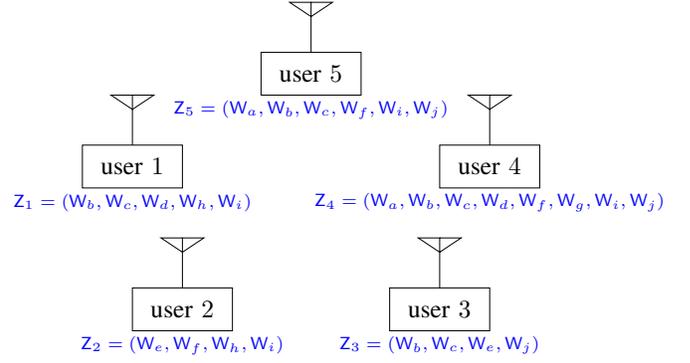

\begin{example} \label{ex:main}
There are five users $V=\Set{1,\dotsc,5}$ in Fig.~\ref{fig:CDESystem}, respectively observing
    \begin{equation}
        \begin{aligned}
            \RZ{1} & = (\RW{b},\RW{c},\RW{d},\RW{h},\RW{i}),   \\
            \RZ{2} & = (\RW{e},\RW{f},\RW{h},\RW{i}),   \\
            \RZ{3} & = (\RW{b},\RW{c},\RW{e},\RW{j}), \\
            \RZ{4} & = (\RW{a},\RW{b},\RW{c},\RW{d},\RW{f},\RW{g},\RW{i},\RW{j}),  \\
            \RZ{5} & = (\RW{a},\RW{b},\RW{c},\RW{f},\RW{i},\RW{j}),
        \end{aligned}  \nonumber
    \end{equation}
with $\RW{j}$ for all $j \in \Set{a,\dotsc,e}$ being an independent uniformly distributed random bit. In CCDE, each $\RW{j}$ represents a packet and the omniscience refers to the recovery of all packets in $\RZ{V}$ by users' broadcasting linear combinations of $\RZ{i}$s over P2P channels \cite{Roua2010}.

By applying the MDA algorithm \cite[Algorithm~1]{Ding2018IT}, we determine the minimum sum-rate $\RCO = \frac{13}{2}$ and an optimal rate vector $(1,\frac{1}{2},\frac{1}{2},\frac{9}{2},0)$, which is an extreme point in $\RRCO(V)$ \cite[Corollary~10]{Ding2018IT}, and also the fundamental partition $\Pat^* = \Set{\Set{1,4,5},\Set{2},\Set{3}}$, which is the finest minimizer of \eqref{eq:Dilworth}. It is not difficult to see that we can improve the fairness of the returned optimal rate vector in $\RRCO(V)$. For example, $(1,\frac{1}{2},\frac{1}{2},4,\frac{1}{2}) \in \RRCO(V)$ is fairer in that user 5 also takes part in the CO instead of being a free rider.
\end{example}

The fairness considered in Example~\ref{ex:main} corresponds to the egalitarian solution \cite{Dutta1989Egalitarian,Dutta1990EgaliConvex}, which tries to make the users have an equal share of the coding rates. The purpose is to motivate them to take part in the CO.
In a system where the users' contribution is unequal, fairness could mean that each user should be penalized proportionally by the coding rates he/she incurs in the CO. In Example~\ref{ex:main}, user $4$ should transmit more since he/she incurs the most coding rates for attaining omniscience, even if the overall coding rates can be distributed to the users more evenly (See Section~\ref{sec:Egalitarian}).
This is another fairness metric called the Shapley value in coalitional game theory. These two fairness metrics are both studied in this paper.

For a fractional rate vector $\rv_V$, if $K \in \ZP$ is the least common multiple (LCM) of all denominators of $r_i$, i.e., $K \rv_V = (K r_i \colon i \in V) \in \ZP^{|V|}$, this rate vector can be implemented by $K$-packet-splitting in CCDE \cite{CourtIT2014,Taj2011,MiloIT2016,Court2011}: dividing each packet into $K$ chunks and letting the users broadcast linear combinations of packet chunks at rate $K\rv_V$. In Example~\ref{ex:main}, both $(1,\frac{1}{2},\frac{1}{2},\frac{9}{2},0)$ and $(1,\frac{1}{2},\frac{1}{2},4,\frac{1}{2})$ can be achieved by $2$-packet-splitting. Therefore, in CCDE, we are also interested in determining a fair fractional optimal rate vector.


\section{Decomposable Coalitional Game}
\label{sec:Game}

We formulate a coalitional game model in this section and show the equivalence of the optimal rate region $\RRCO(V)$ and the core of this game. The purpose is to introduce two game-theoretic solutions, the Shapley value and egalitarian solution in Sections~\ref{sec:Shapley} and \ref{sec:Egalitarian}, respectively, for attaining fairness in $\RRCO(V)$. We also show the decomposition of this game model, a property that will be utilized in Sections~\ref{sec:Shapley} and \ref{sec:Egalitarian} to propose a decomposition method for obtaining the Shapley value and egalitarian solution, respectively.

\subsection{Coalition Game Model}

Let the users in $V$ be self-autonomous decision makers that take part in the CO and assume that, instead of being selfish, they may cooperate with others to form groups. We call $X \subseteq V$ a \textit{coalition} and $V$ the \textit{grand coalition}.
Consider the function $\Fu{\RCO}(X) = H(X) + \RCO - H(V)$.
Here, $\RCO - H(V)$ equals the common randomness $\MI$ in $\RZ{V}$ that is shared by all users in $V$ due to the dual relationship \cite[Theorem~1]{Csiszar2004} \cite{Chan2008tight}
\begin{equation} \label{eq:Dual}
    \RCO = H(V) - \MI.
\end{equation}
Here, $\MI$ is called the multivariate mutual information in \cite{ChanMMI}, or shared information in \cite{Prakash2016}.
Assume that $\MI$ is obtained by a random variable $\RZ{U}$, which does not need to be broadcast over the public channels.
Then, the problem is how to encode the remaining randomness in $\RZ{X}$ given $\RZ{U}$ for all $X \subseteq V$ that is measured by the Dilworth truncation \cite{ChanMMI}
\begin{equation} \label{eq:RemRandom}
    H(X | U) = \FuHat{\RCO}(X) = \min_{\Pat \in \Pi(X)} \sum_{C \in \Pat} \Fu{\RCO}(C).
\end{equation}
We call $\FuHat{\RCO}$ the \textit{characteristic cost} function in that $\FuHat{\RCO}(X)$ specifies the upper bound on the (source) coding cost when the users in $X$ form a coalition so as to jointly encode the randomness in $\RZ{X}$ given $\RZ{U}$. The \emph{coalitional game model} is characterized by the user set $V$ and the characteristic cost function $\FuHat{\RCO}$. We denote it by $\G{V}{\FuHat{\RCO}}$.
In this sense, the game $\G{V}{\FuHat{\RCO}}$ formulates a multi-terminal data compression problem where the users jointly encode the remaining randomness in $\RZ{V}$ that is specified by the set function $\FuHat{\RCO}$.

\begin{example} \label{ex:MutualIndenpendence}
    For the $5$-user system in Example~\ref{ex:main}, the common randomness $\MI = H(V) - \RCO = 10 - \frac{13}{2} = \frac{7}{2}$ is obtained by the random variable $\RZ{U}$. For users 1 and 2, we have
    \begin{equation}
        \begin{aligned}
              H(\Set{1,2} | U)  & = \FuHat{13/2}(\Set{1,2}) \\
                        & = \min \big\{ \Fu{13/2}(\Set{1}) + \Fu{13/2}(\Set{2}),  \Fu{13/2}(\Set{1,2}) \big\}  \\
                                  & = \min \big\{ H(\Set{1}) + H(\Set{2}) - 2H(U), \\
                                  & \qquad H(\Set{1,2}) - H(U) \big\} \\
                                  & = H(\Set{1}) + H(\Set{2}) - 2H(U)  = 2
        \end{aligned}  \nonumber
    \end{equation}
    being the remaining randomness in $\RZ{\Set{1,2}}$ given $U$. The interpretation is that, in order to attain the omniscience with sum-rate $\RCO$, the rate for users 1 and 2 to jointly encode their observations is no more than $2$ bits.
    Or, the maximum cost incurred by users $1$ and $2$ cooperating with each other is $2$ bits of coding rate.
    One can show that \eqref{eq:RemRandom} holds for all $X \subseteq V$.\footnote{An explanation of \eqref{eq:RemRandom} can be found in \cite[Section~IV-B]{ChanMMI}.}
\end{example}

\subsection{Core}
While $\FuHat{\RCO}$ quantifies the maximum coding cost in each coalition, each $\rv_V$ denotes a cost allocation method with each $r_i$ being the source coding rate assigned to user $i \in V$.
The solution set of the game $\G{V}{\Fu{\RCO}}$ is called the \textit{core} \cite{Shoham2008,Shapley1969Core} which contains all $\rv_V$s distributing exactly the total cost $\RCO$ to individual users such that $r(X) \leq \FuHat{\RCO} (X)$ holds for all coalitions $X \subseteq V$.
It is not difficult to see from \eqref{eq:Equivalence} that the core coincides with the optimal rate region $\RRCO(V)$, which is nonempty \cite[Theorem~4]{Ding2018IT}.\footnote{The nonemptiness of the core $\RRCO(V)$ can also be explained by the submodularity of $\FuHat{\RCO}$. See Appendix~\ref{app:game}.}
In the rest of the paper, we will refer to $\RRCO(V)$ as the core or the optimal rate region interchangeably.

The inequality $r(X) \leq \FuHat{\RCO}(X)$ in the core $\RRCO(V)$ also has an interpretation in coalitional game theory.
If a cost allocation method $\rv_V$ results in $r(X) > \FuHat{\RCO}(X)$ for some $X$, the users in $X$ may break the coalition $X$ and seek another $\rv_V$ such that $r(X) \leq \FuHat{\RCO}(X)$. This means the coalition $X$ is not stable.\footnote{This can also be explained by the definition of stability \cite[Section 4.3]{Shapley1971Convex} and the fact that the core is a stable set in \cite[Theorem 8]{Shapley1971Convex}.}
On the other hand, if $r(X) \leq \FuHat{\RCO}(X)$ holds for all $X \subseteq V$, then no user has the incentive to break the coalition $V$ and form a smaller one, i.e., the grand coalition $V$ forms.
In this sense, the core contains all cost allocation methods $\rv_V$ that exactly distribute the sum-cost $r(V) = \RCO$ to all users in a way such that all of them would like to cooperate with others for the purpose of attaining omniscience \cite[Chapter~12]{Shoham2008}.

\subsection{Decomposition}
\label{sec:Decompose}

For any $X, Y \subsetneq V$ such that $X \cap Y = \emptyset$, let $\sqcup$ denote the \textit{disjoint union} and $\rv_X \oplus \rv_Y = \rv_{X \sqcup Y}$ be the \textit{direct sum} of $\rv_X$ and $\rv_Y$.
For example, for $\rv_{\Set{1,3}} = (r_1,r_3) = (3,7)$ and $\rv_{\Set{2,5}} = (r_2,r_5) = (2,4)$, $\rv_{\Set{1,3}} \oplus \rv_{\Set{2,5}} = \rv_{\Set{1,2,3,5}} = (3,2,7,4)$. For $X \subseteq V$, let $\chi_X = (r_i \colon i \in V )$ be the \textit{characteristic vector} of the subset $X$ such that $r_i = 1$ if $i \in X$ and $r_i = 0$ if $i \notin X$. 

For the fundamental partition $\Pat^*$, each $C \in \Pat^*$ defines a subgame $\G{C}{\FuHat{\RCO}}$ with the characteristic cost function $\FuHat{\RCO}(X)$ for all $X \subseteq C$. The core of the subgame $\G{C}{\FuHat{\RCO}}$ is
$$\RRCO(C) = \Set{\rv_C \in P_{C}(\FuHat{\RCO}) \colon r(C) = \FuHat{\RCO}(C)},$$
where the polyhedron $P_{C}(\FuHat{\RCO}) = \Set{\rv_C \in \Real^{|C|} \colon r(X) \leq \FuHat{\RCO}(X),\forall X \subseteq C}$ is a \emph{reduction/projection} of $P(\FuHat{\RCO})$ on to $C$. The following lemma shows the decomposition property of the game $\G{V}{\FuHat{\RCO}}$.

\begin{lemma}[{\cite[Theorem 38 and Lemma 39]{Ding2018IT}}]  \label{lemma:Decompose}
    The game $\G{V}{\FuHat{\alpha}}$ can be decomposed by the fundamental partition $\Pat^*$ so that
    \begin{enumerate}[(a)]
        \item the dimension of $\RRCO(V)$ is $|V| - |\Pat^*|$ and
                \begin{equation}
                    \begin{aligned}
                        \RRCO(V) & = \bigoplus_{C \in \Pat^*} \RRCO(C) \\
                                 & = \Big\{ \bigoplus_{C \in \Pat^*} \rv_C \colon  \rv_C \in \RRCO(C), C \in \Pat^* \Big\}.
                    \end{aligned} \nonumber
                \end{equation}
        \item The following holds for any $\rv_V \in \RRCO(V)$:
              \begin{enumerate}[(i)]
                \item For any $C,C' \in \Pat^*$ such that $C \neq C'$,
                $ \rv_V + \epsilon(\chi_i-\chi_j) \notin \RRCO(V) $, for all $\epsilon>0$, $i \in C$ and $j \in C'$;
                \item For all $C \in \Pat^*$, $ \rv_V + \epsilon(\chi_i-\chi_j) \in \RRCO(V) $ for some $\epsilon>0$ and $i,j \in C$.  \hfill\IEEEQED
              \end{enumerate}

    \end{enumerate}
\end{lemma}

The decomposition of the core $\RRCO(V)$ in Lemma~\ref{lemma:Decompose}(a) interprets the decomposition of the solution set of $\G{V}{\FuHat{\RCO}}$ and the fact that it makes no difference for the users to cooperate in the grand coalition $V$ or in subgames $\G{C}{\FuHat{\RCO}}, \forall C \in \Pat^*$.\footnote{This fact can be seen more clearly via the definition of the decomposable game in Appendix~\ref{app:game}.}
Lemma~\ref{lemma:Decompose}(b) states that the costs, or source coding rates, can be exchanged within a subgame, but not between subgames, which can be explained by the dependence relationship in the remaining randomness as follows.

\subsubsection{Interpretation} \label{sec:DecomposeInterp}

Recall that $\FuHat{\RCO} (X) = H(X|U)$. Due to the fact that $\Pat^*$ is the finest minimizer of \eqref{eq:Dilworth}, we have
\begin{subequations}
    \begin{align}
        & I(C ; C' | U) = \FuHat{\RCO}(C) + \FuHat{\RCO}(C') \nonumber  \\
        & \qquad  - \FuHat{\RCO}(C \sqcup C') = 0, \qquad \forall C,C' \in \Pat^* \colon C \neq C'; \label{eq:ExBetween}\\
        & I(X ; C \setminus X |U) = \FuHat{\RCO}(X) + \FuHat{\RCO}(C \setminus X) \nonumber \\
        & \qquad - \FuHat{\RCO}(C) >  0, \qquad \forall X \subsetneq C.  \label{eq:ExWithin}
    \end{align}
\end{subequations}
Here, \eqref{eq:ExBetween} means that given the common randomness $\MI$ that is obtained by $\RZ{U}$, any two distinct coalitions $C$ and $C'$ in $\Pat^*$ have $\RZ{C}$ and $\RZ{C'}$ mutually independent. That is, to attain the omniscience with the minimum sum-rate $\RCO$, the users in $C$ and $C'$ must encode the exact randomness $H(C|U)$ and $H(C'|U)$, respectively. In other words, the costs or the source coding rates cannot transfer between any two users $i \in C$ and $j \in C'$. This is the interpretation of Lemma~\ref{lemma:Decompose}(b)-(i) and we call it \emph{zero exchange rate} between $i$ and $j$.
On the other hand, \eqref{eq:ExWithin} states that, given the common randomness $\MI$ that is obtained by $\RZ{U}$, any two users $i$ and $j$ in the same coalition $C$ are mutually dependent. In this case, the information amount $I(X ; C \setminus X |U)$ that is mutual to $X$ and $C \setminus X$ can be encoded by either $i \in X$ or $j \in C \setminus X$, i.e., the costs or source coding rates can be transferred between users $i$ and $j$: they have \emph{nonzero exchange rate}.

\begin{figure}[tbp]
	\centering
    \scalebox{0.6}{
%
%
%
\definecolor{mycolor1}{rgb}{0.5,0.5,0.9}%
\definecolor{mycolor2}{rgb}{1,1,0.5}%
\begin{tikzpicture}

\begin{axis}[%
width=3.5in,
height=2.8in,
view={-30}{35},
scale only axis,
xmin=0,
xmax=1.7,
xtick={0,1,2,3},
xlabel={\Large $r_1$},
xmajorgrids,
ymin=0,
ymax=5,
ytick={0,1,2,3,4,5},
ylabel={\Large $r_4$},
ymajorgrids,
zmin=0,
zmax=2.7,
ztick={0,1,2,3,4,5},
zlabel={\Large $r_5$},
zmajorgrids,
axis x line*=bottom,
axis y line*=left,
axis z line*=left,
legend style={at={(0.8,1)},anchor=north west,draw=black,fill=white,legend cell align=left}
]

\addplot3[area legend,solid,fill=mycolor1,draw=black]
table[row sep=crcr]{
x y z\\
1 2 2.5\\
1 4.5 0\\
1.5 4 0\\
1.5 1.5 2.5\\
};
\addlegendentry{\large $\RRCO(\Set{1,4,5})$};

\addplot3[area legend,solid,fill=white!90!black,opacity=4.000000e-01,draw=black]
table[row sep=crcr]{
x y z\\
0 0 0\\
0 4.5 0\\
1 4.5 0\\
1.5 4 0\\
1.5 0 0\\
0 0 0\\
};
\addlegendentry{\large $P_{\Set{1,4,5}}(\Fu{13/2})$};

\addplot3[solid,fill=white!90!black,opacity=4.000000e-01,draw=black,forget plot]
table[row sep=crcr]{
x y z\\
1.5 0 0\\
1.5 4 0\\
1.5 1.5 2.5\\
1.5 0 2.5\\
1.5 0 0\\
};

\addplot3[solid,fill=white!90!black,opacity=4.000000e-01,draw=black,forget plot]
table[row sep=crcr]{
x y z\\
0 4.5 0\\
1 4.5 0\\
1 2 2.5\\
0 2 2.5\\
0 4.5 0\\
};

\addplot3[solid,fill=white!90!black,opacity=4.000000e-01,draw=black,forget plot]
table[row sep=crcr]{
x y z\\
0 0 2.5\\
0 2 2.5\\
1 2 2.5\\
1.5 1.5 2.5\\
1.5 0 2.5\\
0 0 2.5\\
};

\addplot3[solid,fill=white!90!black,opacity=4.000000e-01,draw=black,forget plot]
table[row sep=crcr]{
x y z\\
0 0 0\\
0 0 2.5\\
0 2 2.5\\
0 4.5 0\\
0 0 0\\
};

\addplot3[solid,fill=white!90!black,opacity=4.000000e-01,draw=black,forget plot]
table[row sep=crcr]{
x y z\\
0 0 0\\
1.5 0 0\\
1.5 0 2.5\\
0 0 2.5\\
0 0 0\\
};

\end{axis}
\end{tikzpicture}%
}
	\caption{The core $\RRCO(\Set{1,4,5})$ of the subgame $\G{\Set{1,4,5}}{\FuHat{\RCO}}$ of the $5$-user system in Fig.~\ref{fig:CDESystem}. }
	\label{fig:Core145}
\end{figure}
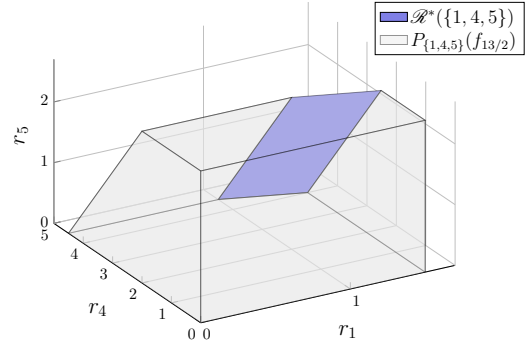

\begin{example} \label{ex:MutualIndenpendence}
    For the $5$-user system in Example~\ref{ex:main}, we have the fundamental partition $\Pat^* = \Set{\Set{1,4,5},\Set{2},\Set{3}}$. The core $\RRCO(V)$ has the dimension of $|V| - |\Pat^*| = 5-3 = 2$ and is decomposed as
    $$ \RRCO(V) = \RRCO(\Set{1,4,5}) \oplus \RRCO(\Set{2}) \oplus \RRCO(\Set{3}) $$
    where $\RRCO(\Set{1,4,5})$, as shown in Fig.~\ref{fig:Core145}, is a 2-dimensional plane and $\RRCO(\Set{2})$ and $\RRCO(\Set{3})$ are singletons containing single points $r_2 = \frac{1}{2}$ and $r_3 = \frac{1}{2}$, respectively.

    Given the common randomness $\MI = H(V) - \RCO = \frac{7}{2}$ that is obtained by $\RZ{U}$, any two distinct $C,C' \in \Pat^*$ are independent, e.g.,
    \begin{equation}
        \begin{aligned}
            I(\Set{1,4,5} ; \Set{2} | U)  = & \FuHat{\RCO}(\Set{1,4,5}) + \FuHat{\RCO}(\Set{2}) \\
                                                 & - \FuHat{\RCO}(\Set{1,2,4,5})= 0;
        \end{aligned} \nonumber
    \end{equation}
    for any $C \in \Pat^*$, any two disjoint $X,Y \subseteq C$ such that $X \sqcup Y = C$ are mutually dependent, e.g.,
    \begin{equation}
        \begin{aligned}
        I(\Set{1,4} ; \Set{5} | U) = & \FuHat{\RCO} (\Set{1,4}) + \FuHat{\RCO} (\Set{5})  \\
                                          & - \FuHat{\RCO} (\Set{1,4,5}) = \frac{5}{2},
        \end{aligned}\nonumber
    \end{equation}
    i.e., in the fundamental partition $\Pat^*$, we have zero exchange rate between coalitions and nonzero exchange rate within a coalition.
\end{example}

The decomposition property in Lemma~\ref{lemma:Decompose} is useful when considering the fairness. Since there is no freedom for the users who belong to distinct coalitions in $\Pat^*$ to negotiate how to allocate coding costs fairly, it suffices to just discuss how to attain fairness within each $C \in \Pat^*$. This will be further summarized in Theorem~\ref{theo:ShapleyDecompose} in Section~\ref{sec:Shapley} and Theorem~\ref{theo:DuttaRayDecompose} in Section~\ref{sec:Egalitarian} that allow distributed computation for attaining the two fair solutions, the Shapley value and egalitarian solution, in the optimal rate region $\RRCO(V)$.

%

\section{Shapley Value}
\label{sec:Shapley}

For an omniscience-achievable rate vector $\rv_V$, it is worth discussing how fairly it can distribute the source coding rates. In the game model $\G{V}{\FuHat{\RCO}}$, fairness is also an important performance metric of a cost allocation method $\rv_V$ in that it promotes the users incentives to cooperate with each other.
In this section, we discuss how to attain fairness by searching the Shapley value in the optimal rate region $\RRCO(V)$.

The Shapley value $\rvS$ is defined in \cite[Theorem 7]{Shapley1953Value} as a unique solution in the core $\RRCO(V)$ with each dimension being
\begin{equation} \label{eq:Shapley}
    \rS_i = \sum_{X \subseteq V \setminus \Set{i}} \frac{|X|! (|V| - |X| - 1)!}{|V|!} \Big( \FuHat{\RCO}(X \sqcup \Set{i}) - \FuHat{\RCO}(X) \Big).
\end{equation}
Here, $\FuHat{\RCO}(X \sqcup \Set{i}) - \FuHat{\RCO}(X) = H(X \sqcup \Set{i} | U) - H(X|U) = H(\Set{i} | X \cup U)$ is the remaining uniqueness in $\RZ{i}$ given the $\RZ{X}$ and the common randomness in $\RZ{U}$. The interpretation is that, to attain the omniscience by the minimum sum-rate $\RCO$, if the users in $X$ encode at the rate $H(X|U)$ first, user $i$ needs to encode at the rate $H(\Set{i} | X \cup U)$.

In the game model $\G{V}{\FuHat{\RCO}}$, $\FuHat{\RCO}(X \sqcup \Set{i}) - \FuHat{\RCO}(X)$ is the marginal coding cost incurred by user $i$ when he/she joins the coalition $X$.
Let $\Phi = (\phi_1,\dotsc,\phi_{|V|})$ such that $\phi_i \in V$ and $\phi_i \neq \phi_j$ for all $i \neq j$ be a \emph{permutation} of $V$. Here, each $\Phi$ denotes the order that the users join the grand coalition $V$, for which, the total cost $\RCO$ can be assigned to individual users by the Edmond greedy algorithm \cite{Edmonds2003Convex}: For $i$ increasing from $1$ to $|V|$, we assign each user the marginal cost
$$ r_i  \coloneqq \FuHat{\RCO}(V_i) - \FuHat{\RCO}(V_{i-1}), $$
where $V_0 = \emptyset$ and $V_i = \Set{\phi_1,\dotsc,\phi_i}$ for all $i \in \Set{1,\dotsc,|V|}$. The resulting $\rv_V$ satisfies $\rv_V \in \RRCO(V)$. The Shapley value $\rvS_V$ is based on the assumption that all the permutations are equiprobable. For each $X \subseteq V \setminus \Set{i} $, user $i$ will be assigned the marginal coding cost $\FuHat{\RCO}(X \sqcup \Set{i}) - \FuHat{\RCO}(X)$ for $|X|! (|V| - |X| - 1)!$ out of $|V|!$ times. Then, $\rvS_V$ assigns each user the expected marginal coding cost he/she incurs over all permutations.

\subsection{Decomposition}
The fairness of $\rvS_V$ can also be explained by its relationship with the extreme points in the core $\RRCO(V)$. Let $\EX(V)$ be the \emph{extreme point set} containing all vertices of the core $\RRCO(V)$.
For a particular permutation $\Phi$, the optimal rate vector returned by the Edmond greedy algorithm is an extreme point of $\RRCO(V)$ and $\EX(V)$ can be constructed by applying the Edmond greedy algorithm for all $|V|!$ permutations of $V$ \cite[Section 3.2]{Fujishige2005}. Based on the definition \eqref{eq:Shapley}, the Shapley value is the mean value of $\EX(V)$ \cite{Shapley1953Value}:\footnote{In this sense, the Shapley value is the gravity center of $\RRCO(V)$ \cite{Shapley1953Value}.}
\begin{equation} \label{eq:ShapleyEX}
    \rvS_V = \frac{\sum_{\rv_V \in \EX(V)} \rv_V}{|\EX(V)|}.
\end{equation}
Since the core $\RRCO(V)$ is decomposed by the fundamental partition $\Pat^*$ (Lemma~\ref{lemma:Decompose}(a)), we have the extreme point set also decomposed as $\EX(V) = \bigoplus_{C \in \Pat^*} \EX(C)$, which leads to the decomposition of the Shapley value in Theorem~\ref{theo:ShapleyDecompose} below.

\begin{theorem} \label{theo:ShapleyDecompose}
    For the Shapley value $\rvS_V$ in the core $\RRCO(V)$, we have
    $$ \rvS_V = \bigoplus_{C \in \Pat^*} \rvS_C, $$
    where $\rvS_C = \frac{\sum_{\rv_C \in \EX(C)} \rv_C}{|\EX(C)|}$ is the Shapley value in the core $\RRCO(C)$ of the subgame $\G{C}{\FuHat{\RCO}}$.
\end{theorem}
\begin{IEEEproof}
For the fundamental partition $\Pat^*$, since $\EX(V) = \bigoplus_{C \in \Pat^*} \EX(C)$, we have
    \begin{equation}
        \begin{aligned}
            \rvS_V & = \frac{\sum_{\rv_V \in \EX(V)} \rv_V }{|\EX(V)|}  \\
                   & = \frac{\sum_{\rv_V \in \bigoplus_{C \in \Pat^*} \EX(C)} \rv_V }{|\bigoplus_{C \in \Pat^*} \EX(C)|} \\
                   & = \frac{\bigoplus_{C \in \Pat^*} \Big( \prod_{C' \in \Pat^* \colon C' \neq C} |\EX(C')| \sum_{\rv_C \in \EX(C)} \rv_C \Big)}{ \prod_{C \in \Pat^*} |\EX(C)| } \\
                   & = \bigoplus_{C \in \Pat^*} \frac{\sum_{\rv_C \in \EX(C)} \rv_C}{|\EX(C)|}  \\
                   & = \bigoplus_{C \in \Pat^*} \rvS_C.
        \end{aligned}  \nonumber
    \end{equation}
Theorem holds.
\end{IEEEproof}

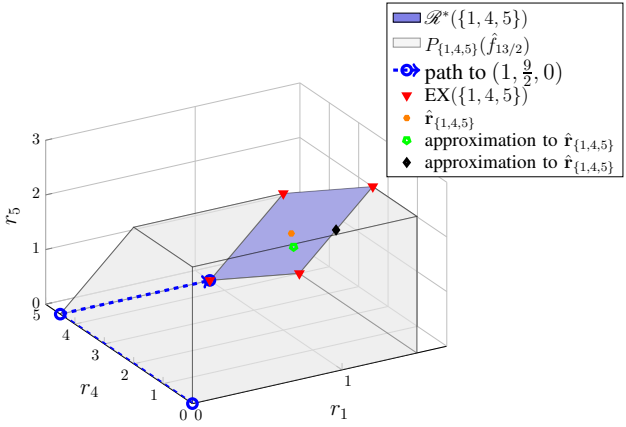
\begin{figure}[tbp]
	\centering
    \scalebox{0.6}{
%
%
%
\definecolor{mycolor1}{rgb}{0.5,0.5,0.9}%
\definecolor{mycolor2}{rgb}{1,1,0.5}%
\begin{tikzpicture}

\begin{axis}[%
width=3.5in,
height=2.8in,
view={-30}{35},
scale only axis,
xmin=0,
xmax=1.7,
xtick={0,1,2,3},
xlabel={\Large $r_1$},
xmajorgrids,
ymin=0,
ymax=5,
ytick={0,1,2,3,4,5},
ylabel={\Large $r_4$},
ymajorgrids,
zmin=0,
zmax=3,
ztick={0,1,2,3,4,5},
zlabel={\Large $r_5$},
zmajorgrids,
axis x line*=bottom,
axis y line*=left,
axis z line*=left,
legend style={at={(0.85,1.25)},anchor=north west,draw=black,fill=white,legend cell align=left}
]

\addplot3[area legend,solid,fill=mycolor1,draw=black]
table[row sep=crcr]{
x y z\\
1 2 2.5\\
1 4.5 0\\
1.5 4 0\\
1.5 1.5 2.5\\
};
\addlegendentry{\large $\RRCO(\Set{1,4,5})$};

\addplot3[area legend,solid,fill=white!90!black,opacity=4.000000e-01,draw=black]
table[row sep=crcr]{
x y z\\
0 0 0\\
0 4.5 0\\
1 4.5 0\\
1.5 4 0\\
1.5 0 0\\
0 0 0\\
};
\addlegendentry{\large $P_{\Set{1,4,5}}(\FuHat{13/2})$};

\addplot3[solid,fill=white!90!black,opacity=4.000000e-01,draw=black,forget plot]
table[row sep=crcr]{
x y z\\
1.5 0 0\\
1.5 4 0\\
1.5 1.5 2.5\\
1.5 0 2.5\\
1.5 0 0\\
};

\addplot3[solid,fill=white!90!black,opacity=4.000000e-01,draw=black,forget plot]
table[row sep=crcr]{
x y z\\
0 4.5 0\\
1 4.5 0\\
1 2 2.5\\
0 2 2.5\\
0 4.5 0\\
};

\addplot3[solid,fill=white!90!black,opacity=4.000000e-01,draw=black,forget plot]
table[row sep=crcr]{
x y z\\
0 0 2.5\\
0 2 2.5\\
1 2 2.5\\
1.5 1.5 2.5\\
1.5 0 2.5\\
0 0 2.5\\
};

\addplot3[solid,fill=white!90!black,opacity=4.000000e-01,draw=black,forget plot]
table[row sep=crcr]{
x y z\\
0 0 0\\
0 0 2.5\\
0 2 2.5\\
0 4.5 0\\
0 0 0\\
};

\addplot3[solid,fill=white!90!black,opacity=4.000000e-01,draw=black,forget plot]
table[row sep=crcr]{
x y z\\
0 0 0\\
1.5 0 0\\
1.5 0 2.5\\
0 0 2.5\\
0 0 0\\
};

\addplot3 [
->,
color=blue,
dashed,
line width=2.0pt,
mark size=3.0pt,
mark=o,
mark options={solid}]
table[row sep=crcr] {
0 0 0\\
0 4.5 0\\
1 4.5 0\\
};
\addlegendentry{\Large path to $(1,\frac{9}{2},0)$ };

\addplot3 [
color=red,
line width=2.0pt,
only marks,
mark=triangle,
mark options={solid,,rotate=180}]
table[row sep=crcr] {
1 2 2.5\\
1 4.5 0\\
1.5 4 0\\
1.5 1.5 2.5\\
};
\addlegendentry{\large $\EX(\Set{1,4,5})$};

\addplot3 [
color=orange,
line width=2.0pt,
only marks,
mark=asterisk,
mark options={solid,,rotate=180}]
table[row sep=crcr] {
1.25 3 1.25\\
};
\addlegendentry{\large $\rvS_{\Set{1,4,5}}$};

\addplot3 [
color=green,
line width=2.0pt,
only marks,
mark=pentagon,
mark options={solid,,rotate=180}]
table[row sep=crcr] {
1.33 3.33 0.833\\
};
\addlegendentry{\large approximation to $\rvS_{\Set{1,4,5}}$};

\addplot3 [
color=black,
line width=2.0pt,
only marks,
mark=diamond,
mark options={solid,,rotate=180}]
table[row sep=crcr] {
1.5 2.75 1.25\\
};
\addlegendentry{\large approximation to $\rvS_{\Set{1,4,5}}$};

\end{axis}
\end{tikzpicture}%
}
	\caption{For the core $\RRCO(\Set{1,4,5})$ of the subgame $\G{\Set{1,4,5}}{\FuHat{\RCO}}$, the extreme point set is $\EX(\Set{1,4,5}) = \Set{ (\frac{3}{2},4,0), (\frac{3}{2},\frac{3}{2},\frac{5}{2}), (1,\frac{9}{2},0), (1,2,\frac{5}{2}) }$, the mean value of which is the Shapley value $\rvS_{\Set{1,4,5}} = (\frac{5}{4},3,\frac{5}{4})$. We apply the random permutation method twice as in Example~\ref{ex:ApprShapley}. We randomly generate $3$ permutations of $1$, $4$ and $5$ each time and get the two approximations of $\rvS_{\Set{1,4,5}}$. In this figure, the path to $(1,\frac{9}{2},0)$ shows an example of how the Edmond algorithm \cite[Algorithm~3]{Ding2018IT} finds the vertex $(1,\frac{9}{2},0)$ corresponding to the permuation $(4,5,1)$. }
	\label{fig:Shapley145}
\end{figure}

\begin{example} \label{ex:Shapley}
    In the core $\RRCO(V)$ of the $5$-user system in Example~\ref{ex:main}, the Shapley value by the definition \eqref{eq:Shapley} is $\rvS_V = (\frac{5}{4},\frac{1}{2},\frac{1}{2},3,\frac{5}{4})$. We have four extreme points in
    \begin{multline}
        \EX(V) = \Big\{ (\frac{3}{2},\frac{1}{2},\frac{1}{2},4,0), (\frac{3}{2},\frac{1}{2},\frac{1}{2},\frac{3}{2},\frac{5}{2}), \\
                        (1,\frac{1}{2},\frac{1}{2},\frac{9}{2},0), (1,\frac{1}{2},\frac{1}{2},2,\frac{5}{2}) \Big\}    \nonumber
    \end{multline}
    such that $\rvS_V = \frac{\sum_{\rv_V \in \EX(V)} \rv_V}{4}$.
    Recall that we have the fundamental partition $\Pat^* = \Set{\Set{1,4,5},\Set{2},\Set{3}}$ that decomposes the game $\G{V}{\FuHat{\RCO}}$ as in Example~\ref{ex:MutualIndenpendence}. According to Theorem~\ref{theo:ShapleyDecompose}, we have
    $$ \rvS_V = \rvS_{\Set{1,4,5}} \oplus \rS_2 \oplus \rS_3, $$
    where $\rvS_{\Set{1,4,5}} = (\frac{5}{4},3,\frac{5}{4}) = \frac{\sum_{\rv_{\Set{1,4,5}} \in \EX(\Set{1,4,5})} \rv_{\Set{1,4,5}}}{4}$ is the Shapley value of the subgame $\G{\Set{1,4,5}}{\FuHat{\RCO}}$ as shown in Fig.~\ref{fig:Shapley145}, $\rS_2 = \frac{1}{2}$ and $\rS_3 = \frac{1}{2}$.
\end{example}

\subsection{Complexity and Approximation}

The complexity of computing the Shapley value is exponentially large in the problem size $|V|$, since the values of $\FuHat{\RCO}(X)$ for all $X \subseteq V$ are required to be calculated to get $\rvS_V$ in \eqref{eq:Shapley}. What makes the situation worse is that determining the value of the Dilworth truncation $\FuHat{\RCO}(X)$ for a given $X$ requires calling SFM algorithms and their complexity is $O(|X| \cdot \SFM(|X|))$. Therefore, it is impractical to obtain the exact value of $\rvS_V$ in large systems.

One alternative approach is to utilize the decomposition property in Theorem~\ref{theo:ShapleyDecompose} to allow distributed and parallel computation. For each coalition $C$ in the fundamental partition $\Pat^*$, let the users in $C$ obtain the Shapley value $\rvS_C$ in the subgame $\G{C}{\FuHat{\RCO}}$ by themselves; All $\rvS_C$ are combined to form the Shapley value $\rvS_V$ of the entire game $\G{V}{\FuHat{\RCO}}$. By doing so, the complexity is determined by the subgame of maximum size $\hat{C} = \argmax\Set{|C| \colon C \in \Pat^*}$. However, the complexity to obtain the Shapley value $\rvS_{\hat{C}}$ in the subgame $\G{\hat{C}}{\FuHat{\RCO}}$ is again exponentially growing in $|\hat{C}|$.

While the high computational complexity is an intrinsic problem of the Shapley value, there are various approximation algorithms proposed in the literature to alleviate this complexity problem.
For example, the random permutation method in \cite{LibenNowell2012} utilizes the fact that the Shapley value is the mean value over the extreme point set in \eqref{eq:ShapleyEX}. The idea is to randomly generate a set of permutations of $V$ of a desired size, e.g., $|V|$ or $|V|^2$ permutations, and apply the Edmond greedy algorithm to determine the corresponding extreme points, the mean of which is an approximation of the Shapley value $\rvS_V$. This approximation method can also be used in combination with the decomposition method in Theorem~\ref{theo:ShapleyDecompose}.

\begin{example} \label{ex:ApprShapley}
    For the $5$-user system in Example~\ref{ex:main}, we first decompose the game into subgames $\G{\Set{1,4,5}}{\FuHat{\RCO}}$, $\G{\Set{2}}{\FuHat{\RCO}}$ and $\G{\Set{3}}{\FuHat{\RCO}}$. For the subgame $\G{\Set{1,4,5}}{\FuHat{\RCO}}$, we randomly select $|\Set{1,4,5}| = 3$ permutations. For example, for $ \Phi = (1,4,5)$, $(1,5,4)$ and $(4,1,5)$, we can generate three extreme points, respectively,
    $$ \Big\{ (\frac{3}{2},4,0),(\frac{3}{2},\frac{3}{2},\frac{5}{2}),(1,\frac{9}{2},0) \Big\} \subsetneq \EX(\Set{1,4,5})$$
    so that the mean value $(\frac{4}{3},\frac{10}{3},\frac{6}{5})$ is an approximation of the Shapley value $\rvS_{\Set{1,4,5}}$ in $\RRCO(\Set{1,4,5})$.
    Note, different permutations might result in different approximations. For example, if we choose three permutations $ \Phi = (1,4,5)$, $(1,5,4)$ and $(5,1,4)$, we would have the approximation being $(\frac{3}{2},\frac{11}{4},\frac{5}{4})$. See the two approximations in Fig.~\ref{fig:Shapley145}.

    By combining the approximation of $\rvS_{\Set{1,4,5}}$ with the ones obtained in other subgames, we have the approximation of the Shapley value $\rvS_V$ of the game $\G{V}{\FuHat{\RCO}}$. For example, the above two approximations generate $(\frac{4}{3},\frac{1}{2},\frac{1}{2},\frac{10}{3},\frac{6}{5})$ and $(\frac{3}{2},\frac{1}{2},\frac{1}{2},\frac{11}{4},\frac{5}{4})$ that are the two approximations to $\rvS_V$.

\end{example}

In Example~\ref{ex:ApprShapley}, we chose no more than $|C|$ permutations for each subgame $C \in \Pat^*$, where the extreme point corresponding to each permutation can be determined by \cite[Algorithm~3]{Ding2018IT}\footnote{The algorithm \cite[Algorithm~3]{Ding2018IT} can be considered as a modified Edmond greedy algorithm. See \cite[Appendix~B]{Ding2018IT} for the explanation. In Fig.~\ref{fig:Shapley145}, the path towards the extreme point $(1,\frac{9}{2},0)$ is generated by \cite[Algorithm~3]{Ding2018IT} for the permutation $\Phi = (4,5,1)$.} in $O(|C| \cdot \SFM(|C|))$ time. Therefore, the overall complexity for approximating the Shapley value $\rvS_V$ is determined by the subgame $\G{\hat{C}}{\FuHat{\RCO}}$ of maximum size as polynomial time $O(|\hat{C}|^2 \cdot \SFM(|\hat{C}|))$.
Accordingly, if we choose $|C|^2$ permutations for each subgame $C \in \Pat^*$, the complexity would be $O(|\hat{C}|^3 \cdot \SFM(|\hat{C}|))$.
We also remark that the approximation algorithm is not unique. In fact, there are many other existing methods, e.g., \cite{LibenNowell2012,Conitzer2004,Fatima2008}, that can be implemented to approximate the Shapley value $\rvS_V$.

%

\section{Egalitarian Solution}
\label{sec:Egalitarian}

The Shapely value $\rvS_V$ is fair in that it penalizes each user based on the expected marginal cost he/she incurs in game $\G{V}{\FuHat{\RCO}}$.
For example, in the $5$-user system in Fig.~\ref{fig:CDESystem}, user $4$ incurs the most expected marginal cost $\FuHat{\RCO}(\Set{4} \sqcup X) - \FuHat{\RCO}(X)$ over all $X \subseteq V \setminus \Set{4}$, the Shapley value $\rvS_V = (\frac{5}{4},\frac{1}{2},\frac{1}{2},3,\frac{5}{4})$ in Example~\ref{ex:Shapley} assigns him/her the most coding cost.

However, this fairness suggested by the Shapley value might not be the desired one in some practical systems. For example, in CCDE where mobile clients are considered as equally privileged peers, it is desirable to find a $\rv_V \in \RRCO(V)$ that allocates the source coding rate as evenly as possible without considering users' prior knowledge of the source $\RZ{V}$. Another example is a WSN containing a large number of battery-powered sensors with equal initial energy budget, where the even allocation of the source coding rate prolongs the overall lifetime of the WSN.
In these cases, it might be more suitable to consider the \emph{egalitarian solution} $\rvE_V$ \cite{Dutta1989Egalitarian,Dutta1990EgaliConvex}, the minimizer of $ \min \Set{ \sum_{i \in V} r_i^2 \colon \rv_V \in \RRCO(V)} $. In this section, we consider a more general quadratic programming \cite{Hokari2002,Hokari2003}
\begin{equation}\label{eq:WeightedQuad}
    \min \Big\{ g(\rv_V) \colon \rv_V \in \RRCO(V) \Big\},
\end{equation}
where $g(\rv_V ) = \sum_{i \in V} \frac{r_i^2}{w_i}$ and $\wv_V \in \RealPP^{|V|}$ is a positive weight vector which could have some interpretations in practical scenarios. For example, $w_i$ could denote the quality of the wireless transmission of user $i$ in CCDE or the remaining battery energy of sensor node $i$ in a WSN.

It is shown \cite{Nagano2012Lex,Nagano2013} that, if the function value of $\FuHat{\RCO}$ can be obtained directly, problem \eqref{eq:WeightedQuad} can be solved in $O(|V| \cdot \SFM(|V|))$ time.\footnote{In the case when $\wv_V = \One = (1,\dotsc,1)$, the minimizer of \eqref{eq:WeightedQuad} is also called the minimum-norm point in $\RRCO(V)$, which can be searched by the algorithm in \cite{Fujishige2011MinNorm} by polynomial time calls of the Dilworth truncation $\FuHat{\RCO}$.}
But, determining the Dilworth truncation $\FuHat{\RCO}(X)$ for a given $X$ has the complexity $O(|X| \cdot \SFM(|X|))$. In addition, the minimizer of \eqref{eq:WeightedQuad} may not be fractional or, if it is fractional, may require splitting each packet into more than $|\Pat^*| - 1$ chunks in CCDE. Since $|\Pat^*| \leq |V|$ and it is shown in \cite[Corollary~28]{Ding2018IT} that there exists an optimal rate vector in $\RRCO(V)$ with an LCM $|\Pat^*| - 1$, it would be of interest to see if we can find a fair optimal rate vector in $\RRCO(V)$ still with LCM $|\Pat^*| - 1$.

\begin{example} \label{ex:Egalitarian}
    Consider the minimizer $\rvE_V$ of \eqref{eq:WeightedQuad} for the $5$-user system in Example~\ref{ex:main}, we have $\rvE_V = (\frac{3}{2},\frac{1}{2},\frac{1}{2},2,2)$ for $\wv_V = \One = (1,\dotsc,1)$ and $\rvE_V = (\frac{3}{2},\frac{1}{2},\frac{1}{2},\frac{12}{5},\frac{8}{5})$ for $\wv_V = (6,1,1,3,2)$. While the former can be implemented by $2$-packet-splitting, the latter requires dividing each packets into $10$ chunks.
\end{example}

In fact, not only the minimizer of \eqref{eq:WeightedQuad}, but also the Shapley value have the problem of incurring more than $(|\Pat^*| - 1)$-packet-splitting. For example, the Shapley value $(\frac{5}{4},\frac{1}{2},\frac{1}{2},3,\frac{5}{4})$ in Example~\ref{ex:Shapley} requires $4$-packet-splitting, where $4 > |\Pat^*| - 1 = 2$, and its approximation $(\frac{4}{3},\frac{1}{2},\frac{1}{2},\frac{10}{3},\frac{6}{5})$ in Example~\ref{ex:ApprShapley} even requires $30$-packet-splitting. Such dividing and reconstructing of packets could be cumbersome or even very impractical.
In the next subsection, we consider how to search for an egalitarian solution in $\RRCO(V)$ that can be implemented by $(|\Pat^*| - 1)$-packet-splitting.

\subsection{Steepest Descent Algorithm}
\label{sec:SDA}

For $K = |\Pat^*| - 1$, let $\Q_{K} = \frac{\Z}{K} $ be the set containing all rational numbers that are divisible by $K$. Consider the problem
\begin{equation} \label{eq:WeightedQuadFrac}
    \min \Big\{ g(\rv_V) \colon \rv_V \in \RRCO(V) \cap \Q_{K}^{|V|}  \Big\}.
\end{equation}
The purpose is to search for a fractional egalitarian solution $\rvE_V$ with an LCM $|\Pat^*| - 1$.
The objective function in \eqref{eq:WeightedQuadFrac} is a separable convex function, for which local optimality w.r.t. the \emph{elementary exchange} $\chi_i - \chi_j$ implies the global optimality. See Lemma~\ref{lemma:LocToGlobMConvex} below.
Here, $\chi_i - \chi_j$ denotes the cost/rate exchange between users $i$ and $j$ in the game $\G{V}{\FuHat{\RCO}}$.\footnote{The optimization criterion in Lemma~\ref{lemma:LocToGlobMConvex} is related to the discrete convexity: The problem in \eqref{eq:WeightedQuad} exhibits $M$-convexity on the real number set \cite[Section 1.4.2]{Murota2003}, which also leads to the $M$-convexity on the fractional number set of \eqref{eq:WeightedQuadFrac}. This is essentially due to the $M$-convexity of a submodular base polyhedron \cite[Theorem 4.12 and Proposition 4.13]{Murota2003}. See also Appendix~\ref{app:Prelim} for the definition of the submodular base polyhedron. }

\begin{lemma}\label{lemma:LocToGlobMConvex}
    In CCDE, $\rvE_V$ is the minimizer of \eqref{eq:WeightedQuadFrac} if and only if, for all $i,j \in V$ and positive integer $\zeta \in \ZPP$ such that $\rv_V^* +  \frac{\zeta}{K}(\chi_i - \chi_j) \in \RRCO(V)$,
    $$ g(\rv_V^*) \leq g \big( \rv_V^* + \frac{\zeta}{K} (\chi_i - \chi_j) \big),$$
    where $K = |\Pat^*| - 1$.
\end{lemma}
\begin{IEEEproof}
    The proof is based on a necessary and sufficient condition for the minimizer of \eqref{eq:WeightedQuad} for any convex function $g$ in \cite[Theorem 20.3]{Fujishige2005}: $\rvE_V$ is the minimizer of \eqref{eq:WeightedQuad} if and only if, for all $i,j \in V$ and positive integer $\epsilon > 0$ such that $\rv_V^* + \epsilon (\chi_i - \chi_j) \in \RRCO(V)$, $ g(\rv_V^*) \leq g(\rv_V^* + \epsilon (\chi_i - \chi_j))$.
    In CCDE, the entropy function $H$ is integer-valued and $\RCO(V)$ is fractional with denominator $K = |\Pat^*| - 1$ so that the value of $\FuHat{\RCO}(X)$ has the denominator $K = |\Pat^*| - 1$ for all $X \subseteq V$. Also, all extreme points in $\EX(V)$ have the LCM $K = |\Pat^*| - 1$ \cite[Corollary~10]{Ding2018IT}. Therefore, for any $\rv_V \in \RRCO(V) \cap \Q_{K}^{|V|}$, if $\rv_V +　\epsilon (\chi_i -\chi_j) \in \RRCO(V)$, then $\rv_V +　\frac{1}{K} (\chi_i -\chi_j) \in \RRCO(V) \cap \Q_{K}^{|V|}$. So, Lemma~\ref{lemma:LocToGlobMConvex} is the result of \cite[Theorem 20.3]{Fujishige2005} on the set $\RRCO(V) \cap \Q_{K}^{|V|}$.
\end{IEEEproof}

	\begin{algorithm} [t]
	\label{algo:SDA}
	\small
	\SetAlgoLined
    \SetKwInOut{Input}{input}\SetKwInOut{Output}{output}
	\SetKwRepeat{Repeat}{repeat}{until}
    \SetKwIF{If}{ElseIf}{Else}{if}{then}{else if}{else}{endif}
    \SetKw{Return}{return}
    \Input{a positive integer $K = |\Pat^*| - 1$ and an initial point $\rv_V^{(0)} \in \RRCO(V) \cap \Q_{|\Pat^*| - 1}^{|V|}$}
	\Output{$\rv_V^{(n)}$, the minimizer of \eqref{eq:WeightedQuadFrac}}
	\BlankLine
	\Begin{
        $n \leftarrow 0$\;
        \Repeat{$\rv_V^{(n+1)} = \rv_V^{(n)}$}{
            \ForAll {$i \in V$}{
                $\dep(\rv_V^{(n)},i) \leftarrow$ the minimal minimizer of
                \begin{equation} \label{eq:MinProbSDA}
                    \min\Set{ \Fu{\RCO}(X) - r^{(n)}(X) \colon i \in X \subseteq V};
                \end{equation}
            }
            $(i^*, j^*) \leftarrow \argmin\Set{g(\rv^{(n)}_V + \frac{1}{K} (\chi_i - \chi_j)) \colon i,j \in V, j \in \dep(\rv_V^{(n)},i) \setminus \Set{i}} $\;
            \eIf{$g(\rv^{(n)}_V + \frac{1}{K} (\chi_{i^*} - \chi_{j^*})) < g(\rv^{(n)}_V )$}{
                $\rv_V^{(n+1)} \leftarrow \rv^{(n)}_V + \frac{1}{K} (\chi_{i^*} - \chi_{j^*})$\;
                $n \leftarrow n +1 $;
            }{
                $\rv_V^{(n+1)} \leftarrow \rv^{(n)}_V$;
            }
        }
        return $\rv_V^{(n)}$\;
    }
	
	\caption{steepest descent algorithm (SDA)}
	\end{algorithm}

Lemma~\ref{lemma:LocToGlobMConvex} directly suggests the steepest descent algorithm (SDA) in Algorithm~\ref{algo:SDA}.\footnote{The SDA algorithm is also based on a discrete convex minimization algorithm in \cite[Section 10.1.1]{Murota2003}. The difference is that we use a dependence function $\dep$ to search the steepest descent direction.}
Also note that, as an input to the SDA, the initial point $\rv_V^{(0)} \in \RRCO(V) \cap \Q_{|\Pat^*| - 1}^{|V|}$ can be searched by the MDA algorithm at the same time when the minimum sum-rate problem is solved \cite[Corollary 28(a)]{Ding2018IT}.
The optimality of the SDA algorithm is stated below.

\begin{theorem} \label{theo:SDA}
    For CCDE, the SDA algorithm in Algorithm~\ref{algo:SDA} generates an estimation sequence $\Set{\rv_V^{(n)}}$ that converges to the minimizer $\rv_V^*$ of \eqref{eq:WeightedQuadFrac}.
\end{theorem}
\begin{IEEEproof}
    Consider the recursive process
    $$ \rv_V^{(n+1)} = \rv_V^{(n)} + \frac{1}{K} (\chi_{i^*} - \chi_{j^*}),$$
    where $(i^*,j^*) = \argmin\Set{ f(\rv_V^{(n)} + \frac{1}{K} (\chi_{i} - \chi_{j})) \colon \rv_V^{(n)} + \frac{1}{K} (\chi_{i} - \chi_{j}) \in \RRCO(V), i,j \in V}$. This is a steepest descent approach: in each iteration $n$, we move from the current estimation $\rv_V^{(n)}$ in the steepest elementary exchange $\chi_{i^*} - \chi_{j^*}$ by a constant step size $\frac{1}{K}$.
    Based on Lemma~\ref{lemma:LocToGlobMConvex}, starting with any initial $\rv_V^{(0)} \in \RRCO(V) \cap \Q_{K}^{|V|}$, the minimum of \eqref{eq:WeightedQuadFrac} is reached when this recursion converges, i.e., when $\rv_V^{(n+1)} = \rv_V^{(n)}$.

    For $\rv_V \in \RRCO(V) \cap \Q_{K}^{|V|}$, consider the \textit{dependence function} \cite[Sections~2.2 and 2.3, Equations~(2.14), (2.15), (2.18) and (2.19)]{Fujishige2005}
        \begin{align}
            &\dep(\rv_V, i)  \nonumber \\
            & \quad = \Set{ j \in V \colon \max\Set{ \epsilon \colon \rv_V + \epsilon (\chi_i - \chi_j) \in \RRCO(V)} > 0 }  \nonumber \\
            & \quad = \bigcap \argmin \Set{ \Fu{\RCO}(X) - r(X) \colon i \in X \subseteq V}. \label{eq:DepAux}
        \end{align}
    The last equality~\eqref{eq:DepAux} states that $\dep(\rv_V, i)$ is the minimal minimizer of $\min \Set{\Fu{\RCO}(X) - r(X) \colon i \in X \subseteq V}$.\footnote{The last equality~\eqref{eq:DepAux} is shown in \cite[Equations~(2.14) and (2.15)]{Fujishige2005} due to the min-max theorem \cite[Corollary~3.4]{Fujishige2005}. The minimizers of $\min \Set{\Fu{\RCO}(X) - r(X) \colon i \in X \subseteq V}$ form a set lattice and the smallest/minimal is the intersection of all minimizers. See \cite[Sections 2.2 and 2.3]{Fujishige2005} for details.}
    A trivial case is that $i \in \dep(\rv_V,i)$. Based on \eqref{eq:DepAux}, we have $ \rv_V +　\frac{1}{K} (\chi_i -\chi_j) \notin \RRCO(V) \cap \Q_{K}^{|V|}$ for all $i, j \in V \colon j \notin \dep(\rv_V,i) \setminus \Set{i}$.
    So, for all iterations $n$ of the recursion above, $\rv_V^{(n)} \in \RRCO(V) \cap \Q_{K}^{|V|}$ and
    \begin{multline}
        (i^*,j^*) = \argmin \big\{ f(\rv_V^{(n)} + \frac{1}{K} (\chi_{i} - \chi_{j})) \colon i,j \in V, \\ j \in \dep(\rv_V,i) \setminus \Set{i} \big\}. \nonumber
    \end{multline}
    Therefore, theorem holds.
\end{IEEEproof}

\begin{remark}
    According to the proofs of Lemma~\ref{lemma:LocToGlobMConvex} and Theorem~\ref{algo:SDA}, if $K \neq |\Pat^*| - 1$, we could have $\rv_V^{(n)} \notin \RRCO(V)$ for some iteration $n$ in the SDA algorithm, or the estimation sequence converges to, but may not reach exactly, the minimizer of \eqref{eq:WeightedQuadFrac}, i.e., the output vector $\rv_{V}^{(n)}$ can be a suboptimal solution of \eqref{eq:WeightedQuadFrac}.
\end{remark}

\begin{example} \label{ex:ConvergeSDA}
   For the $5$-user system in Example~\ref{ex:main}, we first apply the MDA algorithm in \cite{Ding2018IT} and get the minimum sum-rate $\RCO  = \frac{13}{2}$, the fundamental partition $\Pat^* = \Set{\Set{1,4,5},\Set{2},\Set{3}}$ and an extreme point $(1,\frac{1}{2},\frac{1}{2},\frac{9}{2},0) \in \EX(V)$ in the core $\RRCO(V)$.
   By setting $K = |\Pat^*| - 1 = 2$ and $\wv_V = \One$, we start the SDA algorithm with the initial point $\rv_V^{(0)} = (1,\frac{1}{2},\frac{1}{2},\frac{9}{2},0)$.

   At the first iteration $n = 1$, we have
    \begin{equation} \label{eq:DepEx}
        \begin{aligned}
            & \dep(\rv_V^{(0)},1) = \Set{1,4},\ \dep(\rv_V^{(0)},2) = \Set{2},\\
            & \dep(\rv_V^{(0)},3) = \Set{3},\ \dep(\rv_V^{(0)},4) = \Set{4}, \\
            & \dep(\rv_V^{(0)},5) = \Set{4,5}
        \end{aligned}
    \end{equation}
    Then, $\Set{(i,j) \colon j \in \dep(\rv_V^{(0)},i) \setminus \Set{i}} = \Set{(1,4),(4,5)}$. For $\rv_V^{(0)} + \frac{1}{2} (\chi_1 -\chi_4) = (\frac{3}{2},\frac{1}{2},\frac{1}{2},4,0)$ and $\rv_V^{(0)} + \frac{1}{2} (\chi_4 -\chi_5) = (1,\frac{1}{2},\frac{1}{2},4,\frac{1}{2})$, we have $g(\rv_V^{(0)} + \frac{1}{2} (\chi_4 -\chi_5)) < g(\rv_V^{(0)} + \frac{1}{2} (\chi_1 -\chi_4))$ and, therefore, $(i^*,j^*) = (4,5)$. Since $g(\rv_V^{(0)} + \frac{1}{2} (\chi_4 -\chi_5)) < g(\rv_V^{(0)})$, we assign $\rv_V^{(1)} = \rv_V^{(0)} + \frac{1}{2} (\chi_4 -\chi_5) = (1,\frac{1}{2},\frac{1}{2},4,\frac{1}{2})$ and continue the iteration.

    By repeating the same procedure in each iteration, we get the estimation sequence $\Set{\rv_V^{(n)}}$ that results in the update path
    \begin{multline}
        (1,\frac{1}{2},\frac{1}{2},\frac{9}{2},0) \rightarrow (1,\frac{1}{2},\frac{1}{2},4,\frac{1}{2}) \rightarrow (1,\frac{1}{2},\frac{1}{2},\frac{7}{2},1) \\
        \rightarrow (\frac{3}{2},\frac{1}{2},\frac{1}{2},3,1) \rightarrow (\frac{3}{2},\frac{1}{2},\frac{1}{2},\frac{5}{2},\frac{3}{2}) \rightarrow (\frac{3}{2},\frac{1}{2},\frac{1}{2},2,2).  \nonumber
    \end{multline}
    The recursion converges at $n = 6$, where we have $\rv_V^{(6)} = \rv_V^{(5)} = (\frac{3}{2},\frac{1}{2},\frac{1}{2},2,2)$, which is the minimizer $\rvE_V = (\frac{3}{2},\frac{1}{2},\frac{1}{2},2,2)$ of \eqref{eq:WeightedQuadFrac} for $|\Pat^*| - 1 = 2$ and $\wv_V = \One$. Here, $\rvE_V = (\frac{3}{2},\frac{1}{2},\frac{1}{2},2,2)$ is a fractional egalitarian solution, a fair optimal rate vector in $\RRCO(V)$, that can be implemented by $2$-packet-splitting in CCDE.

\end{example}

\subsection{Dependence Function}
\label{sec:dependencefunction}

%
%
Based on \eqref{eq:DepAux}, Lemma~\ref{lemma:Decompose}(b) and the discussion in Section~\ref{sec:Decompose}, it is not difficult to see that, for all $\rv_V \in \RRCO(V)$, if $j \in \dep(\rv_V, i)$ for any $i,j \in V$, then $\RZ{i}$ and $\RZ{j}$ are mutually dependent given the common randomness $\MI = H(V) - \RCO$ obtained by $\RZ{U}$, i.e., $I(\Set{i} ; \Set{j} | U) \neq 0$, hence the name dependence function.
Moreover, due to the fact that $j \in \dep(\rv_V, i)$, we can transfer arbitrarily small, but nonzero, coding cost from user $j$ to user $i$ for encoding the mutually shared information between users $i$ and $j$, which is consistent with the nonzero exchange rate in Section~\ref{sec:Decompose}.

In addition, we must have $\dep(\rv_V, i) \subseteq C$ for the coalition $C \in \Pat^*$ such that $i \in C$, e.g., \eqref{eq:DepEx}. This is because $I(\Set{i} ; \Set{j} | U) = 0$ for all $i \in C, j \in C'$ such that $C \neq C'$ and $I(\Set{i} ; \Set{j} | U) \neq 0$ for all $i,j \in C$, i.e., given the common randomness in $\RZ{U}$, any $\RZ{i}$ is only mutually dependent on any other $\RZ{j}$ in the same coalition $C \in \Pat^*$. This will be formally stated as the decomposition of $\rvE_V$ in Theorem~\ref{theo:DuttaRayDecompose}.

\subsection{Complexity and Distributed Implementation}
\label{sec:SDADistribute}

The SDA algorithm in Algorithm~\ref{algo:SDA} requires oracle calls of $\Fu{\RCO}$, instead of $\FuHat{\RCO}$, which is equivalent to the entry of the entropy function $H$ and avoids the complexity of calculating the Dilworth truncation.
We derive the worst-case complexity of SDA as follows.
For any initial point $\rv_V^{(0)}$, the total number of iterations of the SDA algorithm is $\frac{K \cdot \NormOne{\rv_V^{(0)} - \rvE_V}}{2}$. Let
$$ L(V) = \max \Big\{ \NormOne{\rv_V - \rv'_V} \colon \rv_V, \rv'_V \in \RRCO(V) \cap \Q_{K}^{|V|} \Big\}$$
denote the \emph{$\ell_1$-size} of the core $\RRCO(V)$. The maximum number of iterations of the SDA algorithm is $\frac{K \cdot L(V)}{2}$. The minimization problem \eqref{eq:MinProbSDA} in step 5 in the SDA algorithm is a SFM due to the intersecting submodularity of $\Fu{\RCO}$ \cite[Lemma~3]{Ding2018IT}.
Thus, each iteration of the SDA algorithm completes in $O(|V| \cdot \SFM(|V|))$ time and the overall complexity is $O( K \cdot L(V) \cdot |V| \cdot \SFM(|V|))$.\footnote{The reason that the $\ell_1$-size determines the upper bound on the number of iterations is explained in detail in \cite[Section 10.1.1]{Murota2003}.}

\begin{figure}[tbp]
	\centering
    \scalebox{0.6}{
%
%
\begin{tikzpicture}

\begin{axis}[%
width=4.2in,
height=1.8in,
scale only axis,
xmin=0,
xmax=6,
xlabel={\Large iteration index $n$},
xmajorgrids,
ymin=0,
ymax=5,
ylabel={\Large Error $ \NormOne{\rv_V^{(n)} - \rvE_V}$},
ymajorgrids
]
\addplot [
color=blue,
solid,
mark=asterisk,
mark options={solid},
line width = 1.5
]
table[row sep=crcr]{
0 5\\
1 4\\
2 3\\
3 2\\
4 1\\
5 0\\
6 0\\
};
\end{axis}
\end{tikzpicture}%
}
	\caption{The error measured by the $\ell_1$-norm $\NormOne{\rv_V^{(n)} - \rvE_V}$ of the estimation sequence $\Set{\rv_V^{(n)}}$ generated by the SDA algorithm in Example~\ref{ex:ConvergeSDA} to determine the fractional egalitarian solution in $\RRCO(V)$, the minimizer of $ \min \big\{\sum_{i \in V} r_i^2 \colon \rv_V \in \RRCO(V) \cap \Q_{|\Pat|^* - 1}^{|V|} \big\} $. The error linearly decreases to zero with gradient $-1$, i.e., the $\ell_1$-norm $\NormOne{\rv_V^{(n)} - \rvE_V}$ is reduced by $\frac{2}{|\Pat^*| - 1} = 1$ in each iteration.}
	\label{fig:ConvergeSDA}
\end{figure}
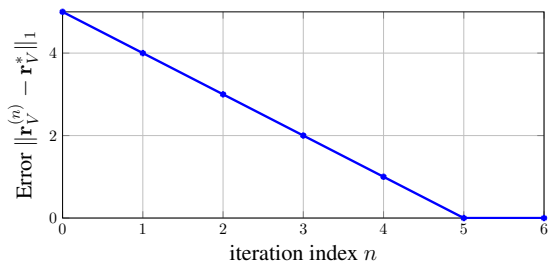

\begin{example} \label{ex:ComplexitySDA}
    For the estimation sequence $\Set{\rv_V^{(n)}}$ generated in Example~\ref{ex:ConvergeSDA} by the SDA algorithm, we show the error of the estimation $\rv_V^{(n)}$ in terms of the $\ell_1$-norm $\NormOne{\rv_V^{(n)} - \rvE_V}$  in \figref{fig:ConvergeSDA}. Since in each iteration of the SDA algorithm, the estimation $\rv_V^{(n)}$ is updated along the steepest elementary exchange $\chi_{i^*} - \chi_{j^*}$ by step size $\frac{1}{K} = \frac{1}{2}$ toward the optimizer $\rvE_V$, we necessarily have $\NormOne{\rv_V^{(n)} - \rvE_V}$ decreased by $\frac{2}{K} = 1$ each time. As in \figref{fig:ConvergeSDA}, we have the error $\NormOne{\rv_V^{(n)} - \rvE_V}$ being a linearly decreasing curve.
    In this case, there are $\frac{K \cdot \NormOne{\rv_V^{(0)} - \rvE_V}}{2} = 5$ iterations in the SDA algorithm so that we incur $5 \cdot |V|$ calls of $O(\SFM(|V|))$.
    In general, since the $\ell_1$-size of $\RRCO(V)$ is $L(V) = 6$, the worst-case complexity of the SDA algorithm when applied to the $5$-user system in Fig.~\ref{fig:CDESystem}, is $6 \cdot |V|$ calls of $O(\SFM(|V|))$.
\end{example}

The SDA algorithm can also be implemented in a decentralized manner: let each user $i$ obtain the dependence function $\dep(\rv_V^{(n)},i)$, a set of mutually dependent users given the common randomness in $\RZ{U}$, by him/herself in steps 4 to 6; the steps 7 to 13 can be completed by users' communications over the broadcast channels. By doing so, the computational complexity incurred at each user is $O( K \cdot L(V) \cdot \SFM(|V|))$.

\subsection{Decomposition}
\label{sec:DecomposeEgalitarian}

Similar to the decomposition of the Shapley value in Theorem~\ref{theo:ShapleyDecompose}, we also have the decomposition property of the egalitarian solution in Theorem~\ref{theo:DuttaRayDecompose}. We omit the proof since it is a direct result of \cite[Corollary~42]{Ding2018IT}, Lemma~\ref{lemma:Decompose} and Lemma~\ref{lemma:LocToGlobMConvex}(b).

\begin{theorem} \label{theo:DuttaRayDecompose}
    For $\rvE_V$ being the egalitarian solution, the minimizer of \eqref{eq:WeightedQuad}, or the fractional egalitarian solution, the minimizer of \eqref{eq:WeightedQuadFrac},
    $$ \rvE_V = \bigoplus_{C \in \Pat^*} \rvE_C, $$
    where $\rvE_C$ is the egalitarian solution or fractional egalitarian solution, respectively, in the core $\RRCO(C)$ of the subgame $\G{C}{\FuHat{\RCO}}$  \hfill\IEEEQED
\end{theorem}

Theorem~\ref{theo:DuttaRayDecompose} states that the egalitarian solution $\rvE_V$ can be determined by allowing the subgames $\G{C}{\FuHat{\RCO}}$ for all $C \in \Pat^*$ to obtain their own $\rvE_C$. This decomposition method can be used in combination with the SDA algorithm so that the complexity is reduced to $O( K \cdot L(\hat{C}) \cdot |\hat{C}| \cdot \SFM(|\hat{C}|))$, where $L(\hat{C})$ is the $\ell_1$-size of the core $\RRCO(\hat{C})$ of the subgame $\G{\hat{C}}{\FuHat{\RCO}}$ of maximum size.
In addition, the users in each subgame can run the SDA algorithm in a distributed manner as discussed in Section~\ref{sec:SDADistribute} and therefore the complexity incurred at each user is $O( K \cdot L(\hat{C}) \cdot \SFM(|\hat{C}|))$.

\begin{remark}
    Theorems~\ref{theo:ShapleyDecompose} and \ref{theo:DuttaRayDecompose} justify the exchange rate resulted from the mutual dependence in Section~\ref{sec:Decompose} when the game $\G{V}{\FuHat{\RCO}}$ is decomposed by the fundamental partition $\Pat^*$ into the subgames $\G{C}{\FuHat{\RCO}}$ for all $C \in \Pat^*$: since the exchange rate, or mutual dependence, is only nonzero inside each subgame $\G{C}{\FuHat{\RCO}}$, we just need to let the users cooperating in the same $\G{C}{\FuHat{\RCO}}$ decide how to attain fairness.
\end{remark}

\begin{figure}[tbp]
	\centering
    \scalebox{0.6}{
%
%
%
\definecolor{mycolor1}{rgb}{0.5,0.5,0.9}%
\definecolor{mycolor2}{rgb}{1,1,0.5}%
\begin{tikzpicture}

\begin{axis}[%
width=3.5in,
height=2.8in,
view={-30}{35},
scale only axis,
xmin=0,
xmax=1.7,
xtick={0,1,2,3},
xlabel={\Large $r_1$},
xmajorgrids,
ymin=0,
ymax=5,
ytick={0,1,2,3,4,5},
ylabel={\Large $r_4$},
ymajorgrids,
zmin=0,
zmax=3,
ztick={0,1,2,3,4,5},
zlabel={\Large $r_5$},
zmajorgrids,
axis x line*=bottom,
axis y line*=left,
axis z line*=left,
legend style={at={(0.85,1.1)},anchor=north west,draw=black,fill=white,legend cell align=left}
]

\addplot3[area legend,solid,fill=mycolor1,draw=black]
table[row sep=crcr]{
x y z\\
1 2 2.5\\
1 4.5 0\\
1.5 4 0\\
1.5 1.5 2.5\\
};
\addlegendentry{\large $\RRCO(\Set{1,4,5})$};

\addplot3[area legend,solid,fill=white!90!black,opacity=4.000000e-01,draw=black]
table[row sep=crcr]{
x y z\\
0 0 0\\
0 4.5 0\\
1 4.5 0\\
1.5 4 0\\
1.5 0 0\\
0 0 0\\
};
\addlegendentry{\large $P(\Fu{13/2}^{\Set{1,4,5}})$};

\addplot3[solid,fill=white!90!black,opacity=4.000000e-01,draw=black,forget plot]
table[row sep=crcr]{
x y z\\
1.5 0 0\\
1.5 4 0\\
1.5 1.5 2.5\\
1.5 0 2.5\\
1.5 0 0\\
};

\addplot3[solid,fill=white!90!black,opacity=4.000000e-01,draw=black,forget plot]
table[row sep=crcr]{
x y z\\
0 4.5 0\\
1 4.5 0\\
1 2 2.5\\
0 2 2.5\\
0 4.5 0\\
};

\addplot3[solid,fill=white!90!black,opacity=4.000000e-01,draw=black,forget plot]
table[row sep=crcr]{
x y z\\
0 0 2.5\\
0 2 2.5\\
1 2 2.5\\
1.5 1.5 2.5\\
1.5 0 2.5\\
0 0 2.5\\
};

\addplot3[solid,fill=white!90!black,opacity=4.000000e-01,draw=black,forget plot]
table[row sep=crcr]{
x y z\\
0 0 0\\
0 0 2.5\\
0 2 2.5\\
0 4.5 0\\
0 0 0\\
};

\addplot3[solid,fill=white!90!black,opacity=4.000000e-01,draw=black,forget plot]
table[row sep=crcr]{
x y z\\
0 0 0\\
1.5 0 0\\
1.5 0 2.5\\
0 0 2.5\\
0 0 0\\
};

\addplot3 [
color=red,
line width=2.0pt,
only marks,
mark=triangle,
mark options={solid,,rotate=180}]
table[row sep=crcr] {
1 2 2.5\\
1 4.5 0\\
1.5 4 0\\
1.5 1.5 2.5\\
};
\addlegendentry{\large $\EX(\Set{1,4,5})$};

\addplot3 [
->,
color=blue,
dotted,
line width=1.5pt,
mark size=1.0pt,
mark=o,
mark options={solid}]
table[row sep=crcr] {
1 4.5 0\\
1 4 0.5\\
1 3.5 1\\
1.5 3 1\\
1.5 2.5 1.5\\
1.5 2 2\\
};
\addlegendentry{\Large path to $\rvE_{\Set{1,4,5}}$ };

\addplot3 [
color=black,
line width=2.0pt,
only marks,
mark=square,
mark options={solid,,rotate=180}]
table[row sep=crcr] {
1.5 2 2\\
};
\addlegendentry{\large $\rvE_{\Set{1,4,5}}$};

\end{axis}
\end{tikzpicture}%
}
	\caption{By applying the SDA algorithm to the subgame $\G{\Set{1,4,5}}{\FuHat{\RCO}}$ of the $5$-user system in Example~\ref{ex:main} with the initial point $\rv_{\Set{1,4,5}}^{(0)} = (1,\frac{9}{2},0)$, we get the estimation sequence $\Set{\rv_{\Set{1,4,5}}^{(n)}}$ resulting an update path toward the fractional egalitarian solution $\rvE_{\Set{1,4,5}}$, the minimizer of $ \min \big\{\sum_{i \in \Set{1,4,5}} r_i^2 \colon \rv_{\Set{1,4,5}} \in \RRCO(\Set{1,4,5}) \cap \Q_{|\Pat^*| - 1}^{3} \big\} $. }
	\label{fig:Egalitarian145}
\end{figure}
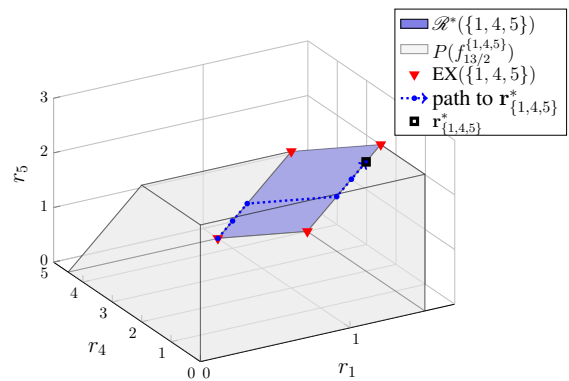

\begin{example}
    For the 5-user system in Example~\ref{ex:main}, consider searching the fractional egalitarian solution w.r.t. $\wv_V = \One$ in Example~\ref{ex:ConvergeSDA} by the decomposition method in Theorem~\ref{theo:DuttaRayDecompose}. We first decompose $\G{V}{\FuHat{\RCO}}$ into subgames $\G{\Set{1,4,5}}{\FuHat{\RCO}}$, $\G{\Set{2}}{\FuHat{\RCO}}$ and $\G{\Set{3}}{\FuHat{\RCO}}$.
For the subgames $\G{\Set{2}}{\FuHat{\RCO}}$ and $\G{\Set{3}}{\FuHat{\RCO}}$, we can directly assign $\rE_2 = \frac{1}{2}$ and $\rE_3 = \frac{1}{2}$, respectively. For the subgame $\G{\Set{1,4,5}}{\FuHat{\RCO}}$, we apply the SDA algorithm and get the following update path to the fractional egalitarian solution $\rvE_{\Set{1,4,5}} = (\frac{3}{2},2,2)$:
    \begin{multline}
        (1,\frac{9}{2},0) \rightarrow (1,4,\frac{1}{2}) \rightarrow (1,\frac{7}{2},1) \\
        \rightarrow (\frac{3}{2},3,1) \rightarrow (\frac{3}{2},\frac{5}{2},\frac{3}{2}) \rightarrow (\frac{3}{2},2,2).  \nonumber
    \end{multline}
    See Fig.~\ref{fig:Egalitarian145}. Then, we get $\rvE_V = \rE_2 \oplus \rE_3 \oplus \rvE_{\Set{1,4,5}} = (\frac{3}{2},\frac{1}{2},\frac{1}{2},2,2)$, the fractional egalitarian solution w.r.t. $\wv_V = \One$ in $\RRCO(V) \cap \Q_{2}^{5}$.

    In this case, we still have $5$ iterations in the SDA algorithm and the convergence performance is exactly the same as in Fig.~\ref{fig:ConvergeSDA}. But, the complexity reduces to $5 \cdot |\Set{1,4,5}|$ calls of $O(\SFM(|\Set{1,4,5}|))$. In general, since $L(\Set{1,4,5}) = 6$, the complexity of the SDA algorithm when applied to the subgame $\G{\Set{1,4,5}}{\FuHat{\RCO}}$, is $6 \cdot |\Set{1,4,5}|$ calls of $O(\SFM(|\Set{1,4,5}|))$.
\end{example}

\section{Conclusion}

We established the equivalence between the optimal rate region of CO and the core of a coalitional game with the characteristic cost function being the Dilworth truncation $\FuHat{\RCO}$ measuring the remaining information $H(X | U)$ in $\RZ{X}$ for all subsets $X \subseteq V$ given the common randomness in $\RZ{U}$.
For attaining fairness in the optimal rate region, we considered the Shapley value and the egalitarian solution. The Shapley value differs from the egalitarian solution in that the fairness is attained if each user $i$ is penalized by the expected marginal cost or source coding rate $H(X \sqcup \Set{i} | U) - H(X | U)$ he/she incurs if in coalition $X$.
By utilizing the fact that the Shapley value is the average over all extreme points in the core, we showed that an approximation, instead of the exact Shapley value, can be obtained by taking the mean over a desired number of randomly generated extreme points.
We also proposed the SDA algorithm for obtaining the egalitarian solution in the core that can be implemented in CCDE by $(|\Pat^*| - 1)$-packet-splitting.
We showed that the game is itself decomposable by the fundamental partition $\Pat^*$ so that, given the common randomness, $\RZ{C}$ and $\RZ{C'}$ for any two distinct $C,C' \in \Pat^*$ are mutually independent, while $\RZ{i}$ and $\RZ{j}$ for all $i, j \in C$ are mutually dependent.
This dependence relationship leads to a decomposition method for obtaining the fair solutions: the Shapley value and the egalitarian solution can be obtained independently within each subgame.

The methods for searching the Shapley value and the egalitarian solution in this paper require the solutions to the minimum sum-rate problem, the value of $\RCO$ and $\Pat^*$ and also an optimal rate vector in $\RRCO(V)$ to initiate the SDA. To further improve the efficiency of attaining fairness in CO, it is worth studying whether we can directly attain the fairness in the optimal rate region without solving the minimum sum-rate problem first.
On the other hand, apart from the fact that the egalitarian solution is more suitable to CCDE and WSN, it is worth understanding to which scenarios the fairness suggested by Shapley value applies.
Finally, the fractional egalitarian solution only determines a fair rate being assigned to each user in CCDE. We still need a complete network coding scheme that also specifies the coefficients in the linear combination of chunks in each transmission.


\appendices

\section{Preliminaries}
\label{app:Prelim}

A set function $f \colon 2^V \mapsto \Real$ is \textit{submodular} if
\begin{equation} \label{eq:SubMIneq}
    f(X) + f(Y) \geq f(X \cap Y) + f(X \cup Y)
\end{equation}
holds for all $X,Y \subseteq V$ \cite[Section 2.3]{Fujishige2005}. A set function $f$ is \emph{intersecting submodular} if the submodular inequality \eqref{eq:SubMIneq} holds for all $X,Y \subseteq V$ such that $X \cap Y \neq \emptyset$\cite[Section 2.3]{Fujishige2005}. $B(f) = \Set{\rv_V \in P(f) \colon \rv_V = f(V)}$ is a \emph{submodular base polyhedron} if $f$ is submodular.
For a submodular function $f \colon 2^V \mapsto \Real$,
\begin{equation} \label{eq:SFM}
    \min \Set{ f(X) \colon X \subseteq V}
\end{equation}
is called \textit{submodular function minimization (SFM)} problem. We assume that the value of $f(X)$ for any $X \subseteq V$ can be obtained by an oracle call and $\delta$ refers to the upper bound on the computation time of this oracle call. It is shown in \cite{Khachiyan1980Ellipsoid,Grotschel2012,Grotschel1981,Iwata2007SFM,IFF2001,Fujishige2011MinNorm} that an SFM problem can be solved in time polynomial in $\delta$. The SFM algorithms proposed in [38]–[43] vary in computation complexity. The exact completion time of an SFM depends on the size of the ground set $V$. For example, the SFM algorithm proposed in \cite{Orlin2009SFM} completes in $O(|V|^5 \cdot \delta + |V|^6)$ time. We denote $O(\SFM(|V|))$ the computation complexity of solving the SFM problem \eqref{eq:SFM}.

A set function $f$ is a \textit{a polymatroid rank function} if it is (a) normalized: $f(\emptyset) = 0 $; (b) monotonic: $f(X) \geq f(Y)$ for all $X,Y \subseteq V$ such that $Y \subseteq X$; and (c) submodular \cite[Section 2.2]{Fujishige2005}. It is shown in \cite[Section 4.2]{FujishigePolyEntropy} that the entropy function $H$ is a polymatroid rank function.

\section{Background on Coalitional Game}
\label{app:game}

Due to the submodularity of $\FuHat{\RCO}$, the game $\G{V}{\FuHat{\RCO}}$ is a convex game, for which the core is always nonempty \cite[Section 2]{Shapley1971Convex}. This also explains the nonemptiness of the core, or the optimal rate region, $\RRCO(V)$. The decomposition property is originally defined for the convex games in \cite{Shapley1971Convex}, which is consistent with the definition of disconnected submodular system in \cite{Fujishige2005,Bilxby1985}.

\begin{definition}[Decomposable Convex Game {\cite[Theorems 3.32 and 3.38, Lemma 3.37]{Fujishige2005}}\footnote{This definition is based on the concept of the separator of a disconnected submodular system in \cite{Fujishige2005,Bilxby1985}.}] \label{def:Deompose}
A convex game $\G{V}{f}$ with the characteristic cost function $f$ is decomposable if
\begin{equation} \label{eq:DecompeEq}
    f(X) = \sum_{C \in \Pat} f(X \cap C), \quad X \subseteq V,
\end{equation}
for some decomposer $\Pat \in \Pi(V)$ such that $\Pat \neq \Set{V}$; Otherwise, $\G{V}{f}$ is indecomposable. For a decomposable convex game $\G{V}{f}$, the subgame $\G{C}{f}$ is convex for each $C \in \Pat$.
\end{definition}

Since \eqref{eq:DecompeEq} always holds for $\Pat = \Set{V}$, an indecomposable game can be considered as convex game with the only decomposer being $\Set{V}$ so that the core $\RRCO(V)$ has the full dimension $|V| - 1$ \cite[Theorem 6(a)]{Shapley1971Convex}.
If a game is decomposable, it must have at least one decomposer other than $\Set{V}$ and all decomposers form a partition lattice, where the finest and coarsest partitions uniquely exist \cite{Narayanan1991PLP,Shapley1971Convex}. 
It is shown in \cite[Theorem~38]{Ding2018IT} that the fundamental partition $\Pat^*$ is the finest decomposer of the game $\G{V}{\FuHat{\RCO}}$.

\section*{Acknowledgment}
The authors would like to thank Prof. Rodney Kennedy for his comments on the early version of the SDA algorithm in \cite{Ding2015ICT}. They are also very grateful to Prof. Chung Chan, Mr. Qiaoqiao Zhou and Prof. Tie Liu for their early contributions to the coalitional game model of CO and the decomposition property of the egalitarian solution in \cite{Ding2015Game,Ding2016ISIT}, as well as the insightful discussions about the interpretation of the Shapley value. The authors would like to thank Dr. Ali Al-Bashabsheh for his comments on the egalitarian solution in Section~\ref{sec:Egalitarian} of this paper.

\bibliographystyle{IEEEtran}
\bibliography{CGBIB}

\end{document}